\documentstyle[12pt,style,k0]{report}
\newcommand{\beq}{\begin{equation}}
\newcommand{\eeq}{\end{equation}}
\newcommand{\bce}{\begin{center}}
\newcommand{\ece}{\end{center}}
\newcommand{\vsp}{\vspace*}
\newcommand{\hsp}{\hspace*}
\newcommand{\s}{\hspace*{0.3em}}
\newcommand{\ben}{\begin{enumerate}}
\newcommand{\een}{\end{enumerate}}
\newcommand{\edoc}{\end{document}}
\newcommand{\btab}{\begin{tabular}[h]}
\newcommand{\etab}{\end{tabular}}
\newcommand{\bfig}{\begin{figure}[h]}
\newcommand{\efig}{\end{figure}}
\newcounter{fig}
\def\fig{\thesection.\thefig}
\newcounter{tab}
\def\tab{\thesection.\thetab}
\input{psfig}

\begin{document}


\titlepage

\centerline{\large Universit\`a degli Studi di Pisa}
\centerline{\large Facolt\`a di Scienze Matematiche, Fisiche e Naturali}
\centerline{\large Corso di Laurea in Fisica}

\vsp{2.5cm}
\centerline{\parbox[t]{14cm}{\Large \bf
            Violazione di CP e rivelazione dei decadimenti
            del $\bf K^0$: studio e simulazione
            del Trigger neutro nell'esperimento NA48.}}

\vsp{2.5cm}
\centerline{\large Tesi di laurea di}
\vsp{0.5cm}
\centerline{\Large \bf Federico Calzolari}

\vsp{2.5cm}
\centerline{{\large Relatore: \hsp{0.5cm} Prof.}
            {\large \bf Giuseppe M. Pierazzini}}

\vsp{5.5cm}
\centerline{------------------------------------}
\centerline{\large Anno Accademico 1994-95}

\clearpage

  \tableofcontents

  \chapter{Introduzione}
\setcounter{fig}{0}
\setcounter{tab}{0}

\noindent
Nell'ambito dell'esperimento NA48 in corso al CERN Super Proton
Syncrothon (SPS) finalizzato a misurare la violazione diretta
di CP in sistemi di particelle $K^0$ attraverso il doppio rapporto
\beq R=\frac{N(K_L\rightarrow\pi^0\pi^0)}{N(K_S\rightarrow\pi^0\pi^0)}
/ \frac{N(K_L\rightarrow\pi^+\pi^-)}{N(K_S\rightarrow\pi^+\pi^-)}
\approx 1-6Re(\frac{\epsilon\prime}{\epsilon}) \eeq
con una precisione di $Re(\frac{\epsilon\prime}{\epsilon})$
superiore a $2 \times 10^{-4}$ \cite{propose},
il lavoro di tesi si propone di effettuare una simulazione
per la rivelazione dei decadimenti neutri  dei $K^0$
ed una successiva analisi dei dati raffrontando l'evento
fisico generato con metodo Montecarlo e quello ricostruito.
Per la misura di $\frac{\epsilon\prime}{\epsilon}$ occorre
isolare tra i vari modi di decadimento dei $K$
quelli   $K_S\rightarrow\pi^0\pi^0$, $K_S\rightarrow\pi^+\pi^-$,
e quelli $K_L\rightarrow\pi^0\pi^0$, $K_L\rightarrow\pi^+\pi^-$.

Due fasci quasi collineari di $K_S$ e $K_L$ sono creati
in modo tale da avere una comune regione di decadimento;
contatori per l'identificazione dei $K_S$ e anticoincidenze
per la soppressione dei decadimenti al di fuori della regione
fiduciale sono disposti lungo la traiettoria dei fasci.
Oltre la zona di decadimento e perpendicolarmente ai fasci
si trovano il calorimetro elettromagnetico a Krypton liquido
per la rivelazione dei fotoni, prodotti del decadimento
di $K^0$ in $2\pi^0 (\rightarrow 4\gamma)$ e
$3\pi^0 (\rightarrow 6\gamma)$, e il calorimetro adronico
per la rivelazione dei decadimenti in $\pi^+\pi^-$.

Il problema principale per la rivelazione dei decadimenti
$K^0\rightarrow 2\pi$ riguarda la e\-li\-mi\-na\-zione del fondo
nei decadimenti $K_L\rightarrow 2\pi$ : questo e' costituito
essenzialmente dai decadimenti
$K_L\rightarrow \pi^\pm   e^\mp \nu$,
$K_L\rightarrow \pi^\pm \mu^\mp \nu$,
$K_L\rightarrow \pi^+   \pi^-   \pi^0 $ per il modo carico,
e dal decadimento
$K_L\rightarrow \pi^0 \pi^0 \pi^0 $ per il modo neutro.
Per ridurre il fondo nei decadimenti carichi sono stati progettati
uno spettrometro magnetico e un contatore di veto per i muoni.
Compito del Trigger neutro e' invece discriminare i
decadimenti $K_S\rightarrow 2\pi^0$ e $K_L\rightarrow 2\pi^0$
da $K_L\rightarrow 3\pi^0$ (questi ultimi costituiscono
piu' del $99\%$ dei decadimenti neutri dei $K_L$),
e arrivare alla ricostruzione di numero di gamma incidenti
sul calorimetro elettromagnetico, energia, baricentro
dell'energia e vertice di decadimento della particella $K$.
Di qui la necessita' di rimuovere quanti piu' possibile
decadimenti $K_L\rightarrow 3\pi^0$, compresi quelli
con soli 4 gamma sul calorimetro elettromagnetico,
cosi' da ridurre il numero di dati da registrare per una
successiva analisi off-line.
Tutto l'apparato di Trigger neutro lavora separatamente
sulle due proiezioni x e y ottenute dalla somma in righe
e colonne delle $128\times 128$ celle del calorimetro
elettromagnetico.

Una vasta collaborazione CERN ha prodotto dati relativi a
decadimenti $K_S\rightarrow 2\pi^0$, $K_L\rightarrow 2\pi^0$,
$K_L\rightarrow 3\pi^0$ generati con metodo Montecarlo.
Il lavoro di tesi parte da una analisi di tali dati,
cui e' stato necessario apportate alcune modifiche,
e procede con:
i)   una simulazione completa dei dispositivi per il trattamento
dei segnali provenienti dalle singole celle del calorimetro
elettromagnetico e per la digitalizzazione dell'informazione,
ii)  una simulazione dell'analisi effettuata dal Trigger neutro,
iii) una ricostruzione dei parametri dell'evento fisico
in ingresso.
La simulazione dei vari componenti e' stata effettuata in
Fortran, al fine di ottimizzare i tempi necessari per
l'analisi numerica di un evento e degli algoritmi impiegati per la
ricostruzione dell'evento stesso.

Attraverso una accurata analisi di decadimenti
$K_S\rightarrow 2\pi^0$, $K_L\rightarrow 2\pi^0$,
$K_L\rightarrow 3\pi^0$, sono state ottenute le distribuzioni
di energia, angolo di apertura e punto di impatto sul
calorimetro elettromagnetico dei gamma prodotti dal decadimento
dei $\pi^0$, e, attraverso la ricostruzione della massa
invariante dei $\pi^0$ a partire dall'energia depositata sulle
singole celle, la traiettoria virtuale dei $\pi^0$. \\
Da una successiva simulazione della ricostruzione dell'evento
fisico da parte della catena elettronica progettata in pipeline,
e' stato possibile ottimizzare le soglie e gli intervalli di
accettazione delle varie quantita' fisiche in gioco, in
corrispondenza di diverse con\-fi\-gu\-ra\-zioni possibili
dell'elettronica stessa.
E' stata infatti testata l'efficienza del Trigger neutro
e degli algoritmi impiegati in condizioni di differenti metodi
di compattazione dell'informazione; tale compattazione si e' resa
necessaria, nonostante una inevitabile perdita in precisione,
per una analisi da parte dell'elettronica sui valori di $rate$
massimi di fun\-zio\-na\-men\-to, previsti intorno a 40 MHz.
Dalla conoscenza dell'evento fisico reale in ingresso
all'apparato, e dalla sua ricostruzione via hardware, sono
state calcolate l'efficienza e la precisione del Trigger neutro,
e raffrontate con le aspettative teoriche. \\
Particolare attenzione e' stata riservata all'analisi del
comportamento del Trigger neutro in condizioni reali di
funzionamento, ovvero con possibili sovrapposizioni spaziali
e temporali, parziali o totali, di eventi, e in presenza
di eventi accidentali, siano essi dovuti a sorgenti esterne,
o a modi di decadimento diversi da quelli analizzati.

Per quanto riguarda la simulazione dell'apparato di
Trigger neutro sono stati creati dei moduli Fortran che
descrivono il comportamento dei vari blocchi: amplificatori
e shaper collegati alle singole celle del calorimetro
elettromagnetico per formare il segnale da analizzare, sommatori
in blocchetti di celle per la compattazione dell'informazione,
digitalizzatori (fast ADC), filtri spaziali e temporali per
l'abilitazione alla somma in righe e colonne dei singoli
blocchetti, Peak-Sum e Peak-Finder progettati all'INFN
sezione di Pisa per il calcolo nelle due proiezioni $x$ e $y$
del numero dei picchi, della temporizzazione fine dei
picchi stessi, e dei momenti dell'energia di ordine 0,1,2
rispetto alla posizione proiettata, necessari alla
Look-up-Table per la ricostruzione di energia totale
depositata sul calorimetro, baricentro dell'energia,
vertice di decadimento del $K^0$.

E' stata dedicata grande attenzione alla verosimiglianza
con la realta' fisica, nonche' alla velocizzazione del sistema
per poter simulare ed analizzare una quantita' di dati
sufficienti a garantire una valida statistica dei valori
inerenti i tre diversi modi di decadimento $K_S\rightarrow 2\pi^0$,
$K_L\rightarrow 2\pi^0$, $K_L\rightarrow 3\pi^0$,
in assenza ed in presenza di eventi accidentali. \\
I risultati di una analisi completa, pur in accordo con le
previsioni teoriche, suggeriscono piccole modifiche al
settaggio di alcuni parametri per una migliore efficienza
nella acquisizione dati.

  \chapter{Violazione di $\bf CP$ e misure sperimentali}

\section{La violazione di $\bf CP$ nei sistemi
         di particelle $\bf K^0$}
\setcounter{fig}{0}
\setcounter{tab}{0}

\noindent
Dalla scoperta della violazione di $CP$ in un sistema di mesoni \k
neutri, nel 1964, da parte di J.H.Christenson, J.Cronin, V.Fitch e
R.Turlay \cite{1964}, svariati esperimenti si sono susseguiti
al fine di una migliore comprensione della portata e dell'origine
del fenomeno.\\
Fino ad allora si era creduto che tutti i processi elementari fossero
invarianti sotto la applicazione di ciascuno dei tre operatori $C$,
$P$, $T$ separatamente.
Tuttavia, gia' nel 1957, una serie di esperimenti mise in evidenza la
violazione di $C$ e $P$ separatamente nei decadimenti deboli dei nuclei
\cite{CPnuc} e delle particelle $\pi$ \cite{CPpi} e $\mu$ \cite{CPmu}.

Il sistema dei mesoni $K^0$ e' finora stato l'unico in cui si sia potuta
misurare con notevole precisone la violazione di $CP$: nel decadimento
dei \kl a lunga vita media in una coppia di \p neutri o carichi,
nell'asimmetria di carica nei decadimenti semileptonici dei \kl.

In un sistema \kz-\akz, ne' \kzv, ne' \akzv sono autostati di $CP$.
Supponendo inizialmente che le interazioni deboli siano assenti, gli
stati \kzv e \akzv sono autostati della Hamiltoniana $H_0$ delle
interazioni forti piu' interazioni elettromagnetiche
\beq H_0=H_f+H_{em} \eeq
\beq H_0|K^0\r=M_0|K^0\r \eeq
dove $M_0$ e' la massa del $K^0$.

Introducendo gli operatori coniugazione di carica $C$ e parita' $P$,
definiamo:
\beq |\ol{K^0}\r=CP|K^0\r \eeq

Dal momento che $H_0$ commuta con $CP$
\beq H_0|\ol{K^0}\r=M_0|\ol{K^0}\r \eeq

{\samepage
\bce \btab{|c|c|} \hline
particella & massa [$GeV$] \\ \hline
\kz        & 0.497         \\
\akz       & 0.497         \\ \hline
\etab \ece
\addtocounter{tab}{1}
\centerline{Tab.\tab: Massa invariante dei \k}}
\vsp{0.5cm}

Inserendo nell'Hamiltoniana $H_0$ le interazioni deboli ($H_d$),
responsabili dei decadimenti di \kzv e \akzv,
\beq H_0=H_f+H_{em}+H_d \eeq
Noto che $H_d$ non conserva ne' $C$, ne' $P$ separatamente, e' comunque
lecito supporre inizialmente che $H_d$ conservi il prodotto $CP$.\\
Gli autostati simultanei di $H_0$ e $CP$ possono essere costruiti
attraverso combinazioni di \kzv e \akzv:
\beq |K_1\r=\frac{1}{\sqrt2}(|K^0\r+|\ol{K^0}\r) \eeq
\beq |K_2\r=\frac{1}{\sqrt2}(|K^0\r-|\ol{K^0}\r) \eeq
lo stato \kuv con $CP=+1$, lo stato \kdv con $CP=-1$.
\kuv e \kdv sono rispettivamente gli stati a vita media breve e lunga.

Considerando i decadimenti del \kz in \dpi o \tpi, poiche' lo stato \dpi
ha $CP=+1$, il decadimento in \dpi e' consentito solo per la particella
\ku, e proibito invece per la particella \kd, che quindi puo' decadere
esclusivamente in \tpi.
Poiche' inoltre il volume nello spazio delle fasi associato al
decadimento di un \kz in \dpi e' molto maggiore di quello associato
ad un decadimento in \tpi, la vita media del \ku e' inferiore a quella
del \kd:
\beq \tau_{K_1} \ll \tau_{K_2} \eeq

{\samepage
\bce \btab{|c|c|} \hline
particella & vita media [$s$]    \\ \hline
\ku        & $0.89\cdot10^{-10}$ \\
\kd        & $5.18\cdot10^{-08}$ \\ \hline
\etab \ece
\addtocounter{tab}{1}
\centerline{Tab.\tab: Vita media dei \k}}
\vsp{0.5cm}

Nel 1964 J.H.Christenson, J.Cronin, V.Fitch e R.Turlay \cite{1964},
utilizzando un apposito spettrometro a due bracci per la rivelazione
dei decadimenti a \dpi munito di rigeneratore, scoprirono che anche il
\kz a lunga vita media (\kd) puo' decadere in \dpi, e piu' precisamente
in \dpic con un branching ratio di 0.002, in \dpiz con un  branching
ratio di 0.001, violando percio' la conservazione di $CP$.

Nel 1965 un esperimento di V.Fitch et al. \cite{1965} dimostro',
attraverso l'interferenza coerente tra le ampiezze di decadimento di \kl
e \ks in \dpic nel vuoto e in presenza di un rigeneratore, che i \dpi
carichi prodotti dal decadimento dei \kl sono identici a quelli prodotti
dal decadimento di \ks, eliminando quindi tutti i tentativi di
spiegazione alternativa che avevano seguito il primo esperimento del
1964, quale tra gli altri la presenza di qualche altra particella a
energia molto bassa.

Gli stati con vita media e massa definita (\kuv e
\kdv nel caso di
conservazione di $CP$) saranno percio' i nuovi stati definiti \ksv e
\klv, rispettivamente a corta vita media e a lunga vita
media.
Utilizzando la notazione introdotta da Wu e Yang \cite{WuYang}, gli
stati \ksv e \klv risultano dalla combinazione di stati \kuv e \kdv:
\beq |K_S\r=\frac{1}{\sqrt2}(|K_1\r+\epsilon|K_2\r) \eeq
\beq |K_L\r=\frac{1}{\sqrt2}(|K_2\r-\epsilon|K_1\r) \eeq
dove $\epsilon$ e' un parametro complesso ($|\epsilon|\approx 2\cdot
10^{-3}$) che tiene conto della violazione di $CP$.

La differenza di massa $M_{K_L}-M_{K_S}=3.521\cdot 10^{-15} GeV$ e'
stata ottenuta attraverso esperimenti di decadimenti in \dpi di miscele
di \ks e \kl.

Data una generica sovrapposizione degli stati \kzv e \akzv, l'evoluzione
temporale del nuovo stato risulta data da:
\beq \Phi(t)=a(t)|K^0\r+b(t)|\ol{K^0}\r=
\left[\begin{array}{c}a(t)\\b(t)\end{array}\right] \eeq
\beq \frac{d}{dt}\Phi(t)=H\Phi(t) \eeq
\beq \frac{d}{dt}\left[\begin{array}{c}a(t)\\b(t)\end{array}\right]=
\frac{1}{i}X\left[\begin{array}{c}a(t)\\b(t)\end{array}\right] \eeq

L'operatore $X$,Hermitiano, puo' essere scomposto in:
\beq X=M-i\frac{\Gamma}{2}\eeq
dove $M$ e $\Gamma$ sono rispettivamente la matrice di massa e la
matrice di decadimento.\\
Gli elementi della matrice $X$ sono:
\beq \l K^0|H|K^0\r \eeq
\beq \l K^0|H|\ol{K^0}\r \eeq
\beq \l\ol{K^0}|H|K^0\r \eeq
\beq \l\ol{K^0}|H|\ol{K^0}\r \eeq

L'invarianza sotto $CPT$ impone che:
\beq X_{11}=X_{22} \eeq
quindi
\beq M_{11}=M_{22} \hsp{1.0cm} \Gamma_{11}=\Gamma_{22} \eeq
mentre per il fatto che l'operatore e' Hermitiano:
\beq X_{12}=X_{21}^* \eeq
quindi
\beq M_{12}=M_{21}^* \hsp{1.0cm} \Gamma_{12}=\Gamma_{21}^* \eeq
Le due matrici $M$ e $\Gamma$ risultano quindi essere:
\beq M=\left[\begin{array}{cc}M_0&M_{12}\\
M_{12}^*&M_0\end{array}\right]\eeq
\beq \Gamma=\left[\begin{array}{cc}\Gamma_0&\Gamma_{12}\\
\Gamma_{12}^*&\Gamma_0\end{array}\right]\eeq
con $M_0$ e $\Gamma_0$ reali.

L'equazione agli autovalori per l'operatore $X$ ha come soluzioni:
\beq M_S-\frac{i}{2}\Gamma_S=X_{11}+\sqrt{X_{12}X_{21}} \eeq
\beq M_L-\frac{i}{2}\Gamma_L=X_{11}-\sqrt{X_{12}X_{21}} \eeq
relative agli autostati
\beq |K_S\r=\frac{1}{\sqrt{1+|\epsilon|^2}}(|K_1\r+\epsilon|K_2\r) \eeq
\beq |K_L\r=\frac{1}{\sqrt{1+|\epsilon|^2}}(|K_2\r+\epsilon|K_1\r) \eeq
dove e' stato introdotto il parametro $\epsilon$
\beq \epsilon=\frac{\sqrt{X_{12}}-\sqrt{X_{21}}}
{\sqrt{X_{12}}+\sqrt{X_{21}}} \eeq

Sostituendo a $X_{ij}$ i valori ricavati dall'equazione agli autovalori,
si ottiene:
\beq \epsilon=\frac{\frac{1}{2}\Im\Gamma_{12}+i\Im M_{12}}
{\frac{i}{2}(\Gamma_S-\Gamma_L)-(M_S-M_L)} \eeq
Utilizzando i dati sperimentali per $M_S$, $M_L$, $\Gamma_S$,
$\Gamma_L$, il valore del modulo di $\epsilon$ risulta essere:
\beq |\epsilon|=(2.3\pm0.02)\cdot10^{-3} \eeq
trascurando $\Im\Gamma_{12}$, la fase di $\epsilon$ e' determinata
dal denominatore, ed e' pari a:
\beq arg(\epsilon)=tg^{-1}2\frac{M_L-M_S}{\Gamma_S-\Gamma_L}
\approx 45^\circ\pm1.5^\circ \eeq

Il parametro $\epsilon$ tiene conto della violazione indiretta di $CP$,
attraverso il decadimento di una miscela di stati di opposta $CP$
(\kuv e \kdv).

Nei decadimenti di \kz in \dpi lo stato finale puo' essere
caratterizzato dal numero quantico di isospin $I=0$, oppure $I=2$,
entrambe combinazioni simmetriche di due stati a isospin $I=1$.
Indicando con $|0\r$ e $|2\r$ gli stati di isospin corrispondenti
rispettivamente a $I=0$ e $I=2$ per i decadimenti in \dpi, si ricavano
quattro ampiezze di transizione:
\beq \l0|T|K_S\r \eeq
\beq \l2|T|K_S\r \eeq
\beq \l0|T|K_L\r \eeq
\beq \l2|T|K_L\r \eeq
dove T e' l'Hamiltoniana responsabile dei decadimenti che violano $CP$,
riducibili a tre quantita' normalizzando rispetto all'ampiezza
$\l0|T|K_S\r$:
\beq \epsilon=\frac{\l0|T|K_L\r}{\l0|T|K_S\r} \eeq
\beq \epsilon\prime=\frac{\l2|T|K_L\r}{\l0|T|K_S\r} \eeq
\beq \omega=\frac{\l2|T|K_S\r}{\l0|T|K_S\r} \eeq

Le quantita' osservabili, disponibili da esperimenti, sono:
\beq \eta_{+-}=\frac{\l\pi^+\pi^-|T|K_L\r}{\l\pi^+\pi^-|T|K_S\r}
=\frac{\epsilon+\epsilon\prime}{1+\frac{\omega}{\sqrt2}} \eeq
\beq \eta_{00}=\frac{\l\pi^0\pi^0|T|K_L\r}{\l\pi^0\pi^0|T|K_S\r}
=\frac{\epsilon-2\epsilon\prime}{1-\sqrt2\omega} \eeq
\beq \delta_L=\frac{(K_L\ra\pi^+\ell^-\nu)-(K_L\ra\pi^-\ell^+\nu)}
{(K_L\ra\pi^+\ell^-\nu)+(K_L\ra\pi^-\ell^+\nu)} \eeq
dove $\ell$ e' un leptone: elettrone(positrone) $e^\mp$ o muone
$\mu^\pm$.

Definendo gli stati a due pioni \p in funzione degli autostati
dell'operatore isospin:
\beq \l\pi^+\pi^-|=\sqrt{\frac{2}{3}}\l0|+\sqrt{\frac{1}{3}}\l2| \eeq
\beq \l\pi^0\pi^0|=\sqrt{\frac{1}{3}}\l0|-\sqrt{\frac{2}{3}}\l2| \eeq
poiche' $\omega\ll1$ :
\beq \eta_{+-}=\epsilon+\epsilon\prime \eeq
\beq \eta_{00}=\epsilon-2\epsilon\prime \eeq
\beq \delta_L=2\Re\epsilon \eeq

La grandezza fisica misurabile $\delta_L$ e' riferita ad una asimmetria
di carica nei decadimenti semileptonici $K_L\ra\pi^\pm\ell^\mp\nu$,
ed evidenzia anch'essa la violazione di $CP$ nei sistemi di \kz.

Introducendo le ampiezze di transizione di \kz e \akz nei canali a
isospin $I=0$ e $I=2$:
\beq \l0|T|K^0\r     =A_0   e^{i\delta_0} \eeq
\beq \l2|T|\ol{K^0}\r=A_0^* e^{i\delta_0} \eeq
\beq \l0|T|K^0\r     =A_2   e^{i\delta_2} \eeq
\beq \l2|T|\ol{K^0}\r=A_2^* e^{i\delta_2} \eeq
il parametro $\epsilon\prime$ risulta essere
\beq
\epsilon\prime=\frac{1}{\sqrt2}\Im(\frac{A_2}{A_0})
e^{i(\delta_2-\delta_0)} \eeq
ed e' direttamente collegabile alla violazione diretta di $CP$,
attraverso il decadimento diretto di uno stato a $CP$ definita in uno
stato finale a $CP$ opposta.

La scomposizione delle ampiezze di decadimento dei $K^0$ osservabili in
$\epsilon$ e $\epsilon\prime$ corrisponde ad una separazione degli
effetti di violazione di $CP$ a livello della matrice di massa $M$
($\epsilon$) e a livello della matrice di decadimento $\Gamma$
($\epsilon\prime$).

Il doppio rapporto
\beq R=\left|\frac{\eta_{00}}{\eta_{+-}}\right|^2=
\frac{N(K_L\ra\pi^0\pi^0)}{N(K_S\ra\pi^0\pi^0)}/
\frac{N(K_L\ra\pi^+\pi^-)}{N(K_S\ra\pi^+\pi^-)}
\approx 1-6\Re(\frac{\epsilon\prime}{\epsilon}) \eeq
costituisce la misura cui e' finalizzato l'esperimento NA48 al CERN
Super Proton Synchroton (SPS).

\section{La misura di $\bf R$}
\setcounter{fig}{0}
\setcounter{tab}{0}

\noindent
Il modulo e la fase di $\eta_{+-}$ sono stati misurati in maniera
precisa attraverso fasci di \ks e \kl preparati attraverso un
rigeneratore:
\beq \eta_{+-}=[(2.274\pm0.0022)\cdot10^{-3}]
\s\mbox{exp}[i(0.778\pm0.021)] \eeq

Per quanto riguarda la misura di modulo e fase di $\eta_{00}$, questa
risulta piu' complessa da determinare, e comporta errori sistematici
maggiori che nella misura di $\eta_{+-}$; questo perche' dopo il
decadimento del \k in \dpiz, il \piz decade con vita media brevissima
(dell'ordine di $10^{-16}\s s$) in due fotoni, per cui ne risulta un evento
a quattro fotoni, energia e posizione di ciascuno dei quali devono
essere misurate con estrema precisione.\\

{\samepage
\bce \btab{|c|c|c|} \hline
particella & vita media [$s$] & massa [$GeV$] \\ \hline
\pip       & $10^{-08}$       & $0.139$       \\
\pim       & $10^{-08}$       & $0.139$       \\
\piz       & $10^{-16}$       & $0.135$       \\ \hline
\etab \ece
\addtocounter{tab}{1}
\centerline{Tab.\tab: Vita media e massa dei $\pi$}}
\vsp{0.5cm}

\section{Misure di ${\frac{\epsilon\prime}{\epsilon}}$}
\setcounter{fig}{0}
\setcounter{tab}{0}

\noindent
La prima prova della violazione diretta di $CP$ e' stata annunciata
della collaborazione NA31 del CERN di Ginevra nel 1988 \cite{B317},
con un valore
\beq \Re(\frac{\epsilon\prime}{\epsilon})
=(2.30\pm0.65)\cdot10^{-3} \eeq

Nel 1990 l'esperimento E731 al Fermilab di Chicago \cite{E731} ha
riportato il valore
\beq \Re(\frac{\epsilon\prime}{\epsilon})
=(0.74\pm0.59)\cdot10^{-3} \eeq

L'attuale migliore stima di $\frac{\epsilon\prime}{\epsilon}$ dovrebbe
quindi risultare dalla media, pesata rispetto al numero di decadimenti
analizzati, di queste due precedenti misurazioni:
\beq \Re(\frac{\epsilon\prime}{\epsilon})
=(2.10\pm0.90)\cdot10^{-3} \eeq

Nell'esperimento NA31, concettualmente simile a NA48, salvo una minore
precisione, la precisione statistica e' stata limitata dal numero
relativamente basso di decadimenti \kldpiz.\\
Altri esperimenti sono stati effettuati negli ultimi anni, senza pero'
fornire misure con precisione tale da poter essere confrontate con
queste.

\section{Violazione di $\bf CP$ in altri sistemi fisici}
\setcounter{fig}{0}
\setcounter{tab}{0}

Sistemi analoghi al sistema \kz-\akz sono $D^0$-$\ol{D^0}$ e
$B^0$-$\ol{B^0}$, cui e' possibile applicare lo stesso formalismo
impiegato per i mesoni \k neutri.
Tali sistemi sono pero' costituiti da particelle molto piu' pesanti,
con vite medie molto piu' brevi, e con un vasto numero di possibili
stati finali di decadimento.

Nonostante in questi sistemi la violazione di $CP$ a livello della
matrice di massa $M$ sia maggiore che nel caso del sistema \kz-\akz,
risulta molto piu' difficile da misurare.
Nel sistema \kz-\akz infatti, essendo la massa del \kz poco piu' grande
della massa totale dei \tpi, il che implica una lunga vita media del
\kl, esiste una grande differenza tra la vita media delle particelle \ks
e \kl, consentendo quindi una buona identificazione, all'atto
dell'esperimento, delle particelle stesse.

  \chapter{L'apparato sperimentale NA48}
\setcounter{fig}{0}
\setcounter{tab}{0}

\noindent
L'esperimento NA48 \cite{propose}, attualmente in corso al CERN Super
Proton Synchroton (SPS), consta di un apparato di rivelazione per la
misura della violazione diretta di $CP$ nei decadimenti dei \k neutri,
attraverso il doppio rapporto
\beq R=\frac{N(K_L\rightarrow\pi^0\pi^0)}{N(K_S\rightarrow\pi^0\pi^0)}
/ \frac{N(K_L\rightarrow\pi^+\pi^-)}{N(K_S\rightarrow\pi^+\pi^-)}
\approx 1-6Re(\frac{\epsilon\prime}{\epsilon}) \eeq
con una precisione di $Re(\frac{\epsilon\prime}{\epsilon})$
superiore a $2 \times 10^{-4}$.\\
La misura di $\frac{\epsilon\prime}{\epsilon}$ e' ottenuta attraverso un
raffronto dei rates relativi di decadimento di \ks e \kl in \dpic e
\dpiz.\\

\bce
\vsp{7cm}
\nopagebreak
\addtocounter{fig}{1}
Fig.\fig: Apparato sperimentale NA48
\ece
\vsp{0.5cm}

Per una misura precisa occorre che le efficienze di rivelazione per il
modo carico e per il modo neutro risultino il piu' possibile uguali;
questo e' ottenuto attraverso l'utilizzo degli stessi contatori per la
rivelazione dei decadimenti in \dpic e \dpiz, siano essi derivati da
\ks, sia da \kl.\\

{\samepage
In tabella sono riportati i principali modi di decadimento dei \ks e
\kl \cite{K0decay}:

\def\e{\cdot10^}
\bce \btab{|c|l|l|} \hline
particella & decadimento              & branching ratio \\ \hline
$K_S$      & $\pi^+\pi^-            $ & $0.686      $   \\
           & $\pi^0\pi^0            $ & $0.314      $   \\
           & $\pi^+\pi^-\gamma      $ & $0.002      $   \\
           & $\pi^0\pi^0\pi^0       $ & $0.005\e{-2}$   \\
           & $\pi^+\pi^-\pi^0       $ & $0.004\e{-2}$   \\
           & $e^+e^-                $ & $0.001\e{-2}$   \\
           & $\gamma\gamma          $ & $0.002\e{-3}$   \\
           & $\mu^+\mu^-            $ & $0.003\e{-4}$   \\ \hline
$K_L$      & $\pi^\pm  e^\mp\nu     $ & $0.386      $   \\
           & $\pi^\pm\mu^\mp\nu     $ & $0.270      $   \\
           & $\pi^0\pi^0\pi^0       $ & $0.217      $   \\
           & $\pi^+\pi^-\pi^0       $ & $0.124      $   \\
           & $\pi^\pm e^\mp\nu\gamma$ & $0.013      $   \\
           & $\pi^+\pi^-            $ & $0.002      $   \\
           & $\pi^0\pi^0            $ & $0.009\e{-1}$   \\
           & $\gamma\gamma          $ & $0.005\e{-1}$   \\
           & $\pi^0\gamma\gamma     $ & $0.002\e{-1}$   \\
           & $\pi^0\pi\pm e^\mp\nu  $ & $0.006\e{-2}$   \\
           & $\pi^+\pi^-\gamma      $ & $0.004\e{-2}$   \\ \hline
\etab \ece
\addtocounter{tab}{1}
\centerline{Tab.\tab: Modi di decadimento dei \k neutri}}
\vsp{0.5cm}

Il problema principale riguardante la rivelazione dei decadimenti dei
\kz consiste nella eliminazione del fondo per i decadimenti \kldpi.
Questo e' costituito essenzialmente dai decadimenti
$K_L\ra\pi^\pm e^\mp\nu$, $K_L\ra\pi^\pm\mu^\mp\nu$,
$K_L\ra\pi^+\pi^-\pi^0$ per il modo carico, e dal decadimento
$K_L\ra\pi^0\pi^0\pi^0$ per il modo neutro.

Uno spettrometro magnetico e' progettato per ridurre il fondo nei
decadimenti carichi, ed un calorimetro elettromagnetico a Krypton
liquido per ridurre il fondo nei decadimenti neutri.

\section{I fasci di $\bf K_S$ e $\bf K_L$}
\setcounter{fig}{0}
\setcounter{tab}{0}

\noindent
Disponendo di due fasci quasi collineari di \ks e \kl aventi una
comune regione di decadimento, all'interno di una zona di $120\s m$
di lunghezza e $2.5\s m$ di diametro, mantenuta sotto vuoto spinto
(Vacuum tank), e' possibile confrontare direttamente i rate relativi
dei decadimenti di \ks e \kl in \dpic e \dpiz.
L'angolo di convergenza dei due fasci di \ks e \kl e' inferiore alla
divergenza del singolo fascio, al fine di eliminare tutte le possibili
fonti di differenza sistematiche nella rivelazione dei modi di
decadimento di \ks e \kl.

Un fascio di protoni primari con intensita' di $10^{12}\s protoni/s$
ed energia di $450\s GeV$ viene utilizzato per produrre il fascio di \kl
dall'impatto su una targhetta \kl.
I \kl cosi' creati passano attraverso un primo collimatore ed un primo
contatore, mentre i protoni e le altre particelle cariche, frutto
dell'interazione dei protoni con la targhetta \kl, vengono deviati dalla
direzione del fascio \kl.\\
Dopo un secondo collimatore, una piccola frazione dei protoni primari di
intensita' pari a $3\cdot10^{07}\s protoni/s$ viene nuovamente devialta
verso il fascio dei \kl e, attraverso un deflettore magnetico, viene
convogliata lungo la beam-line costituita dal fascio \kl.
Un ultimo deflettore devia nuovamente il fascio di protoni e lo
indirizza, attraverso un contatore, verso la targhetta \ks per la
generazione dei \ks.

La differenza angolare nella traiettoria del fascio di \ks rispetto al
fascio di \kl e' di $0.6\s mrad$.\\
In questo modo il numero di decadimenti \ksdpi risulta essere
confrontabile col numero di decadimenti \kldpi all'interno della comune
regione di decadimento.

La distinzione tra decadimenti di \ks e decadimenti di \kl e' affidata
ad un sistema di tagging-counter per l'identificazione e l'etichettatura
dei protoni impiegati per la ge\-ne\-ra\-zio\-ne dei \ks.
Questa identificazione viene effettuata misurando la differenza
temporale tra il passaggio dei protoni nei contatori posti a monte della
targhetta \ks e il tempo dell'evento nel rivelatore.
Gli eventi per cui questa differenza temporale risulta essere
all'interno di un intervallo di accettazione vengono etichettati come
\ks, mentre gli altri vengono identificati come \kl.

Qualunque inefficienza nel sistema di conteggio dei \ks implica
quindi che un decadimento \ks venga identificato come un decadimento
\kl; d'altra parte, ogni conteggio accidentale nel contatore \ks fa si'
che un decadimento \kl venga identificato come un decadimento \ks.\\
Importante e' mantenere basse queste probabilita' di transizione, e
questo avviene attraverso la scelta di una finestra temporale di
accettazione di $5\s ns$.
Cosi' facendo l'errore nell'identificazione di \ks e \kl e' ridotto a
circa il $5\%$.

Il rate totale degli eventi nel rivelatore e' previsto intorno a
$10^6\s Hz$.

\section{Le anticoincidenze e i contatori di veto}
\setcounter{fig}{0}
\setcounter{tab}{0}

\noindent
I decadimenti \kltpiz avvengono con un rate 200 volte maggiore dei
decadimenti \kldpiz utili alla misura di $R$.
E' percio' essenziale che una fase di pre-trigger fornisca un veto per
la maggior parte di tale fondo, al fine di bloccare l'acquisizione dati
per questi eventi.
Parte di questo compito e' affidato ad una serie di anticoincidenze,
costituite da un doppio strato di scinitillatore preceduto e intercalato
da strati di Piombo per la rivelazione dei gamma, e posizionate in
maniera tale da coprire un ampio angolo solido esterno al rivelatore.
Il cono di accettanza dell'apparato risulta cosi' delimitato da una
serie di sette anticoincidenze disposte ad anello intorno alla
beam-line, in grado di rivelare i gamma con traiettorie esterne ai
rivelatori.\\
L'efficienza di tali anticoincidenze per i decadimenti \kltpiz e' pari
a 0.75 per un'energia del \kl di $100\s GeV$, e sale fino a 1.00 per
\kl ad energia inferiore a $50\s GeV$, che risultano quindi avere un
angolo di decadimento dei \piz , quindi dei gamma, maggiore.
In questo modo viene soppressa parte parte del fondo per idecadimenti
neutri in \tpiz.

Per il modo carico e' necessario sopprimere il conteggio per i
decadimenti $K_L\ra\pi^\pm e^\mp\nu$, $K_L\ra\pi^\pm\mu^\mp\nu$,
$K_L\ra\pi^+\pi^-\pi^0$.
I decadimenti $K_L\ra\pi^+\pi^-\pi^0$ vengono anticoincisi attraverso la
rivelazione dei \dpic sul calorimetro adronico e del \piz sul
calorimetro elettromagnetico;
i decadimenti $K_L\ra\pi^\pm e^\mp\nu$  vengono anticoincisi dalla
ricostruzione sul calorimetro adronico della massa invariante
di un solo \piz;
i decadimenti $K_L\ra\pi^\pm\mu^\mp\nu$  vengono anticoincisi attraverso
l'utilizzo di quattro larghi piani di scintillatori utilizzati come
contatori di veto per i muoni $\mu^\pm$.

\section{Il calorimetro elettromagnetico}
\setcounter{fig}{0}
\setcounter{tab}{0}

\noindent
Per la rivelazione dei decadimenti neutri del \kz e' necessario disporre
di un rivelatore rispondente alle sequenti caratteristiche:
\begin{itemize}
\item possibilita' di registrare eventi a piu' fotoni ad un rate di
      $1\s MHz$
\item risoluzione in energia $\sigma(E)=0.035\sqrt{E}$
\item risoluzione spaziale inferiore a $1\s mm$
\item risoluzione temporale inferiore a $1\s ns$
\end{itemize}
Per soddisfare tali richieste e' stato progettato un calorimetro
elettromagnetico ottagonale di $256\s cm$ di raggio ad alte efficienza e
risoluzione.

Aspetti peculiari di tale apparato di rivelazione sono:
\begin{itemize}
\item segmentazione del rivelatore in una matrice di $128\times128$
      celle di $2\s cm\times2\s cm$ ciascuna
\item bassa capacita' dei singoli canali del rivelatore, per permettere
      una veloce lettura dell'energia depositata
\item utilizzo di Krypton liquido ad alta densita' in una struttura
      quasi omogenea
\end{itemize}

Ogni singola cella ha una superficie utile di $2\times2\s cm^2$, ed e'
costituita da un elettrodo centrale ad alta tensione inserito tra due
elettrodi di massa distanti $1\s cm$ da esso e condivisi da due celle
adiacenti sul piano orizzontale.\\
Gli elettrodi, di una lega di Rame e Berillio, sono divisi in strisce
la cui altezza ($2\s cm$) definisce la struttura verticale della singola
cella.
Tali strisce, lunghe $125\s cm$, sono state disposte a zig-zag con
angoli di $\pm50\s mrad$ dall'asse della beam-line, al fine di evitare
che un gamma incidente sullo spessore ($40\s \mu m$)della singola
striscia di rame possa incanalarvisi senza interagire col mezzo
rivelatore, e quindi non rilasciare energia in nessuna cella.

Come mezzo ionizzabile viene impiegato Krypton liquido (LKr) per una
profondita' pari a quella degli elettrodi di rivelazione ($125\s cm$),
corrispondente a circa $27$ lunghezze di radiazione.

All'interno della singola cella ogni drift-gap e' di $1\s cm$,
corrispondente ad un drift-time di $2.8\s \mu s$ per una tensione
applicata di $5000\s V$ \cite{LKr}.

Uno sciame elettromagnetico, prodotto dall'interazione di un gamma con
la materia (Krypton liquido), che si sviluppa all'interno del
calorimetro, vi deposita una certa quantita' di carica che induce un
impulso di corrente in corrispondenza di ciascuna cella del
rivelatore.\\
Tale impulso, di forma triangolare, viene raccolto da amplificatori di
carica, direttamente connessi ad ogni singola cella, e contenuti
all'interno del criostato per il mantenimento del Krypton allo stato
liquido.\\
Il segnale lascia poi il criostato ed e' filtrato attraverso uno
shaping-amplifier, che provvede a formare un impulso in uscita di
ampiezza proporzionale alla derivata del segnale in ingresso, e
caratterizzato da una durata di $110\s ns$, con larghezza a meta'
altezza di $60\s ns$, ed undershoot del $3\%$ della durata di
$2.5\s \mu s$.
La raccolta di un'energia pari a $1\s GeV$ sulla singola cella fornisce
una corrente con un'ampiezza massima di $3\s \mu A$, ed una successiva
ampiezza del segnale di shaper di $10\s mV$.
L'energia massima depositabile sulla singola cella e' di
circa $50\s GeV$.

La risoluzione in energia e' data da \cite{LKr}:
\beq (\frac{\sigma_E}{E})^2=(\frac{0.035}{\sqrt{E}})^2+0.005^2 \eeq
e il profilo dell'impulso di corrente in funzione del punto di impatto
del gamma sulla singola cella, a parita' di energia depositata, risulta
essere quasi piatto.

La risoluzione spaziale e' inferiore a $2\s mm$, ed e' data da:
\beq \sigma_x=\frac{4.2}{\sqrt{E\s[GeV]}}+0.6\s mm \eeq
\beq \sigma_y=\frac{4.3}{\sqrt{E\s[GeV]}}+0.6\s mm \eeq

\section{L'odoscopio neutro}
\setcounter{fig}{0}
\setcounter{tab}{0}

\noindent
Oltre al calorimetro elettromagnetico, un odoscopio neutro a
scintillazione ha il compito di rivelare il tempo esatto di arrivo
dell'evento, cosi' da costituire una fase di pre-trigger per i
decadimenti carichi.

Tale odoscopio,diviso in due semipiani, richiede una coincidenza
destra/sinistra per l'accettazione degli eventi.

\section{Lo spettrometro magnetico}
\setcounter{fig}{0}
\setcounter{tab}{0}

\noindent
I momenti delle particelle cariche vengono misurati in uno spettrometro
magnetico costituito da due serie di quattro camere a drift, tra le
quali e' interposto un dipolo magnetico centrale.\\
Con un campo magnetico integrato equivalente ad una variazione di
momento trasverso di $200\s MeV/c$, la risoluzione nella misura del
momento di una particella e' limitata dallo scattering multiplo sul
materiale costituente le camere a:
\beq \frac{\Delta p}{p}=0.006 \eeq
Questo e' in accordo con la scelta di camere a drift aventi una
risoluzione spaziale di $100\s \mu m$.

Ciascuna delle quattro camere a drift consiste di quattro coppie di
piani di fili formanti due sistemi di assi cartesiani ruotati fra loro
di $45^\circ$ per una migliore precisione nella misura.
Ciascun piano e' costituito da 240 fili distanti fra loro $1\s cm$.

Le camere a drift lavorano con una miscela $70/30$ di Argon-isoButano, e
la velocita' di drift risulta di $5\s cm/\mu s$.

\section{Il calorimetro adronico}
\setcounter{fig}{0}
\setcounter{tab}{0}

\noindent
Oltre lo spettrometro magnetico e' disposto un calorimetro adronico ad
Argon liquido per la rivelazione della posizione dei \pip e \pim che,
avendo vita media piu' lunga dei \piz, arrivano pressoche' tutti a
colpire i rivelatori senza decadere.\\

{\samepage
\bce \btab{|c|c|} \hline
particella & vita media [$s$] \\ \hline
\pip       & $10^{-08}$       \\
\pim       & $10^{-08}$       \\
\piz       & $10^{-16}$       \\ \hline
\etab \ece
\addtocounter{tab}{1}
\centerline{Tab.\tab: Vita media dei $\pi$}}
\vsp{0.5cm}

Dal punto di impatto del $\pi^\pm$ sul calorimetro adronico e dalle
informazioni ricevute dalle camere a drift, e' possibile ricostruire
l'esatta traiettoria del $\pi^\pm$, e, dal raggio di curvatura impresso
dal campo magnetico $B$ dello spettrometro magnetico, l'impulso della
particella.

\section{L'odoscopio carico}
\setcounter{fig}{0}
\setcounter{tab}{0}

\noindent
Una fase di pre-trigger per i decadimenti carichi viene effettuata da un
odoscopio posto di fronte al calorimetro elettromagnetico.
Tale odoscopio, diviso in quattro quadranti, richiede una coincidenza in
due quadranti opposti per l'accettazione dell'evento

\section{Il contatore muonico}
\setcounter{fig}{0}
\setcounter{tab}{0}

\noindent
Uno scintillatore costituito da 3 piani paralleli, posto oltre il
calorimetro elettromagnetico e il calorimetro adronico, ha il compito di
rivelare i decadimenti di \ks e \kl in cui sia presente un mesone $\mu$,
nonche' la posizione del punto di impatto del $\mu$ stesso, al fine di
porre un veto sull'accettazione di tali eventi.

\section{Trigger}
\setcounter{fig}{0}
\setcounter{tab}{0}

\noindent
Mentre i \ks decadono quasi esclusivamente in \dpic e \dpiz, per quanto
riguarda i \kl, questi decadono in \dpi soltanto nello $0.3\s \%$ dei
casi, e risultano pertanto coperti dal fondo degli altri predominanti
modi di decadimento.

Un complesso sistema di trigger a piu' livelli \cite{NET} e' necessario
ai fini dell'esperimento per la rivelazione dei decadimenti in \dpi,
siano essi dovuti al decadimento di un \ks, o di un \kl.
Tenuto conto dell'alta intensita' del fascio, al trigger e' richiesto di
avere una buona efficienza a rates elevati, e di riuscire a eliminare
una elevata percentuale del fondo.

I candidati per i decadimenti in \dpiz sono selezionati tra quegli
eventi con una energia ricostruita tra $60\s GeV$ e $180\s GeV$, con un
vertice ricostruito tra $2.1\s m$ e $48.9\s m$ dalla posizione
dell'ultimo collimatore, e un numero di gamma sul calorimetro
elettromagnetico compreso tra $3$ e $4$.

I candidati per i decadimenti in \dpic sono selezionati tra i vari
eventi utilizzando gli stessi tagli in energia, vertice e
numero di gamma adottati per i decadimenti neutri \cite{B317}.

\section{Una prima fase di trigger}
\setcounter{fig}{0}
\setcounter{tab}{0}

\noindent
Una fase di pre-trigger viene effettuata, al fine di ridurre il rate
degli evnti da analizzare di circa 10 volte, dal Trigger di livello 1.
Elaborando in tempo rapido le informazioni provenienti da
anticoincidenze, contatori di veto, odoscopio carico ed odoscopio
neutro, tale trigger abilita o meno la successiva analisi da parte dei
trigger di livello 2: il Trigger carico e il Trigger neutro.

\section{Il Trigger carico}
\setcounter{fig}{0}
\setcounter{tab}{0}

\noindent
Il trigger carico, utilizzato per distinguere i decadimenti \dpic dal
fondo, utilizza, per la ricostruzione dell'evento, le informazioni
provenienti dagli odoscopi carichi, dal calorimetro adronico, e dallo
spettrometro magnetico.\\
Per i decadiment carichi e' richiesta nei rivelatori una coincidenza
simmetrica rispetto alla beam-line.
Dalle informazioni del calorimetro adronico e dello spettrometro
magnetico, il trigger carico ricostruisce energia, massa invariante,
momento trasverso e vertice di decadimento del \k.

\section{Il Trigger neutro}
\setcounter{fig}{0}
\setcounter{tab}{0}

\noindent
Compito principale del Trigger neutro e' la rimozione del fondo
dominante costituito dai decadimenti \kltpiz nei decadimenti neutri.

I \piz decadono con vita media brevissima generando ciascuno due gamma,
che vengono poi rivelati sul calorimetro elettromagnetico.
I decadimenti \kltpiz si presentano come eventi a 6 gamma; tali eventi
possono essere rivelati sul calorimetro come decadimenti a 6 gamma nel
caso in cui tutti i fotoni raggiungano il rivelatore, come decadimenti
con numero di gamma inferiore a 6 nel caso in cui alcuni fotoni seguano
traiettorie esterne all'apparato di rivelazione o vengano persi in
quanto poco energetici.

I tagli principali che consentono una notevole riduzione del fondo
\kltpiz vengono effettuati sul numero di picchi, sull'energia totale
rilasciata nel calorimetro elettro\-ma\-gne\-tico, sul baricentro
dell'energia, sul vertice di decadimento del \k.
Al Trigger neutro e' richiesto di lavorare in modo conservativo rispetto
al numero di eventi, ovvero senza scartare eventuali eventi sporcati da
un evento accidentale poco energetico, che potrebbero  magari essere
ricostruiti ugualmente bene attraverso una successiva analisi off-line
degli eventi registrati.

Il Trigger neutro lavora non sulle informazioni provenienti da tutte le
$13500$ celle del calorimetro elettromagnetico, bensi' sulle proiezioni
$x$ e $y$ ottenute dalla somma in righe e colonne, ulteriormente
accorpate in gruppi di due, dell'energia depositata sulle singole
celle.\\
All'interno di ogni singola riga (colonna) le celle vengono sommate in
blocchetti $8\times2$ ($2\times8$) e quindi digitalizzate.
I blocchetti cosi' creati vengono analizzati dal filtro temporale, il
quale abilita alla somma in righe e colonne il singolo blocchetto,
quello temporalmente seguente e i due precedenti, nel caso in cui il
blocchetto in esame risulti maggiore di una soglia selezionabile
dall'esterno.

I vantaggi di una siffatta compattazione dell'informazione, nonostante
una inevitabile perdita in precisione ed efficienza, consistono nella
possibilita' da parte dell'apparato elettronico di un'analisi sui valori
di rate massimi di funzionamento, previsti intorno a $40\s MHz$.\\
Per una maggiore velocita' di analisi, l'elettronica e' stata progettata
in pipeline, con tempi di elaborazione e di transito dell'informazione
nel singolo stadio ampiamente inferiori ai $25\s ns$, al fine di
garantire una stabilita' dei registri al momento della commutazione del
clock.

Dispositivi principali dell'apparato di Trigger neutro sono il Peak-Sum
e il Peak-Finder, progettati all'INFN sezione di Pisa per il calcolo
nelle due proiezioni $x$ e $y$ del numero di picchi in energia presenti
sul calorimetro elettromagnetico, della temporizzazione fine dei picchi
stessi, e dei momenti dell'energia di ordine 0,1,2 rispetto alla
posizione proiettata, necessari ai moduli Look-up Table per la
ricostruzione di energia, baricentro dell'energia, vertice di
decadimento del \k.

\section{Il Trigger Supervisor}
\setcounter{fig}{0}
\setcounter{tab}{0}

\noindent
Per gli evnti giudicati buoni dal trigger carico e dal Trigger neutro,
una successiva e piu' accurata analisi, resa possibile dai tagli
effettuati e quindi da un rate di eventi decisamente inferiore, mira ad
una migliore e piu' precisa ricostruzione delle quantita' fisiche
osservabili, al fine di attivare l'acquisizione dati su disco per gli
eventi candidati ad essere frutto di decadimenti \ksdpi e \kldpi.

Oltre alle quantita' precedentemente calcolate, ovvero numero di picchi,
energia, baricentro dell'energia, vertice di decadimento del \k, il
trigger supervisor provvede ad una accurata ricostruzione della massa
invariante della particella originaria, e puo' quindi porre nuovi tagli
su questa quantita', riducendo cosi' ulteriormente il fondo dei
decadimenti del \k a tre corpi.

Il Trigger Supervisor provvede anche ad un raffronto incrociato delle
informazioni provenienti dal trigger carico e dal Trigger neutro per
una analisi dei decadimenti misti (in particelle cariche e neutre
insieme).\\

{\samepage
Alla fine dell'analisi effettuata on-line dall'apparato sperimentale, il
fondo previsto per i quattro modi di decadimento utili alla misura e'
\cite{B317}:
\def\e{\cdot10^}
\bce \btab{|c|c|} \hline
decadimento        & fondo residuo (\%) \\ \hline
$K_S\ra\pi^+\pi^-$ & $0.03$             \\
$K_S\ra\pi^0\pi^0$ & $0.07$             \\
$K_L\ra\pi^+\pi^-$ & $0.63$             \\
$K_L\ra\pi^0\pi^0$ & $2.67$             \\ \hline
\etab \ece
\addtocounter{tab}{1}
\nopagebreak
\centerline{Tab.\tab: Fondo residuo}}
\vsp{0.5cm}

Il fondo residuo maggiore risulta essere, come previsto, per il modo di
decadimento \kldpiz.

  \chapter{Simulazione Montecarlo}

\section{Decadimenti del $\bf K$ e rivelatori: Montecarlo}
\setcounter{fig}{0}
\setcounter{tab}{0}

\noindent
Al fine di studiare i vari aspetti del nuovo esperimento NA48 EPSI
inerente la violazione di $CP$ nei decadimenti dei \k neutri, e' stato
creato da una vasta collaborazione CERN un programma di simulazione per
generare in tempi relativamente brevi una vasta gamma di decadimenti con
metodo Montecarlo.\\
Il programma, denominato appunto New Montecarlo (NMC) \cite{NMC}, genera
eventi di decadimento di \ks e \kl partendo da una banca dati (particle
bank), relativa alla particella o\-ri\-gi\-na\-ria, dove sono contenute
informazioni quali tipo di particella, quadriimpulso, vertice di
decadimento, energia, vita media.

NMC simula i processi di decadimento di tali particelle e la risposta
dei rivelatori utilizzati per le misure.
Attraverso banche dati contenenti informazioni relative a traiettorie e
modi di decadiemnto delle varie particelle, vengono ricostruiti i
percorsi delle particelle e dei loro prodotti di decadimento fino ai
rivelatori, disposti lungo la traiettoria del fascio e
perpendicolarmente ad esso.

Dopo il decadimento dei \piz in gamma nella camera a vuoto per il modo
di decadimento neutro, viene simulata la rivelazione dei gamma da parte
del calorimetro elettromagnetico a Krypton liquido attraverso l'utilizzo
di librerie di sciami elettromagnetici, prodotti dall'interazione dei
gamma con la materia.
Tali librerie di sciami, prodotte con programmi di simulazione del CERN
NASIM-GEANT, vengono impiegate da NMC per simulare l'energia depositata
celle per cella sul calorimetro elettromagnetico.

\clearpage

\section{Simulazione fisica}
\setcounter{fig}{0}
\setcounter{tab}{0}

\noindent

\subsection{I fasci $\bf K_S$ e $\bf K_L$}

\noindent
I \ks e \kl vengono generati utilizzando lo spettro di produzione
\beq \frac{d^2N}{dp\s d\Omega}=\frac{np^2}{4p_0}
[1.30 e^{-(8.5\frac{p}{p_0}+3.0\frac{p^2}{\theta^2})}
+4.35 e^{-(13.0\frac{p}{p_0}+3.5\frac{p^2}{\theta^2})}] \eeq
dove
\beq \theta_S=4.2\s mrad \hsp{1.0cm} \mbox{relativo a \ks} \eeq
\beq \theta_S=2.4\s mrad \hsp{1.0cm} \mbox{relativo a \kl} \eeq
\beq p_0=450\s \frac{GeV}{c} \eeq

e una distribuzione di decadimento
\beq d(z)=\frac{1}{\lambda} e^{\frac{z}{\lambda}}\eeq
dove
\beq \lambda=\gamma c \tau \eeq

Lo spettro di decadimento e' ottenuto pesando lo spettro di produzione
dei \k con la distribuzione di decadimento e integrando sul volume
fiduciale nel quale i decadimenti sono generati.

Le particelle \ks e \kl vengono generate entro un intervallo di energia
ed entro un intervallo di vertice di decadimento tipici
dell'esperimento:\\

{\samepage
\bce \btab{|c|r|}  \hline
Energia min. & $60\s GeV$  \\
Energia max. & $180\s GeV$ \\   \hline
Vertice min. & $-100\s cm$ \\
Vertice max. & $3500\s cm$ \\ \hline
\etab \ece
\addtocounter{tab}{1}
\centerline{Tab.\tab: Limiti valori generati}}
\vsp{0.5cm}

Il vertice di decadimento e' definito come distanza del punto di
decadimento del \k dalla targhetta \ks.\\
Circa $6\s m$ oltre la targhetta \ks e' collocata l'ultima serie di
collimatori di diametro $0.3\s cm$ e $3.0\s cm$ rispettivamente per \ks
e \kl.
E' percio' possibile che i gamma prodotti da un \k decaduto poco
prima del collimatore finale riescano comunque a giungere sul
rivelatore attraverso il foro di apertura del collimatore stesso, e
vengano quindi rivelati sul calorimetro elettromagnetico.
Per quanto riguarda i \ks un taglio netto sul limite inferiore del
vertice di decadimento e' effettuato, oltre che dall'ultima serie di
collimatori, dall'anticoincidenza \ks impiegata per porre un veto
sull'accettazione di quegli eventi con fotoni esterni all'apparato di
rivelazione.\\
I decadimenti di \kl distanti oltre $38\s m$ dalla targhetta \ks non
sono presi in consi\-de\-ra\-zio\-ne, in quanto l'angolo di apertura
dei gamma generati da tali decadimenti non consente ai gamma di
giungere oltre il foro del calorimetro elettromagnetico.

Il fascio di \ks, distante sul piano perpendicolare alla beam-line di
$7.2\s cm$, all'altezza della targhetta \ks, dal fascio \kl, viene
deflesso, attraverso il collimatore \ks, verso i rivelatori con un
angolo di $0.6\s mrad$ dal fascio \kl.
La divergenza dei fasci \ks e \kl, seppur minima, e' tenuta in
considerazione dal programma di simulazione.

\subsection{I modi di decadimento dei $\bf K$}

\noindent
Il programma Montecarlo e' in grado di simulare, attraverso banche dati
dedicate, 15 diversi modi di decadimento del \k, oltre che 3 differenti
modi di decadimento del \p.

I modi di decadimento di \ks e \kl:\\
\bce \btab{|c|l|} \hline
particella & decadimento                     \\ \hline
$K_S$      & $K_S\ra \pi^+\pi^-            $ \\
           & $K_S\ra \pi^0\pi^0            $ \\ \hline
$K_L$      & $K_L\ra \pi^\pm  e^\mp\nu     $ \\
           & $K_L\ra \pi^\pm\mu^\mp\nu     $ \\
           & $K_L\ra \pi^0\pi^0\pi^0       $ \\
           & $K_L\ra \pi^+\pi^-\pi^0       $ \\
           & $K_L\ra \pi^\pm e^\mp\nu\gamma$ \\
           & $K_L\ra \pi^+\pi^-            $ \\
           & $K_L\ra \pi^0\pi^0            $ \\
           & $K_L\ra \mu^+\mu^-\gamma      $ \\
           & $K_L\ra e^+ e^-\gamma         $ \\
           & $K_L\ra \pi^+\pi^- e^+ e^-    $ \\
           & $K_L\ra \pi^0 e^+ e^-         $ \\
           & $K_L\ra \pi^0\pi^0\gamma      $ \\
           & $K_L\ra e^+ e^- e^+ e^-       $ \\ \hline
\etab \ece
\addtocounter{tab}{1}
\centerline{Tab.\tab: Modi di decadimento NMC \ks, \kl}
\vsp{0.5cm}

\clearpage

I modi di decadimento di \p:\\
\bce \btab{|c|l|} \hline
particella & decadimento                     \\ \hline
$\pi$      & $\pi^\pm\ra \mu^\pm\nu        $ \\
           & $\pi^0  \ra \gamma\gamma      $ \\
           & $\pi^0  \ra e^+ e^-\gamma     $ \\ \hline
\etab \ece
\addtocounter{tab}{1}
\centerline{Tab.\tab: Modi di decadimento NMC \p}
\vsp{0.5cm}

\section{I rivelatori}
\setcounter{fig}{0}
\setcounter{tab}{0}

\subsection{Il sistema di etichettatura dei $\bf K_S$}

\noindent
Se la particella viene creata da un protone primario incidente sulla
targhetta \ks, il sistema etichetta tale evento come \ks.

\subsection{Le anticoincidenze e i contatori di veto}

\noindent
Le anticoincidenze per i \ks e \kl, utilizzate per sopprimere i
decadimenti con gamma al di fuori della traiettoria dei rivelatori, sono
costituite da contatori a scintillazione preceduti da un convertitore
metallico di Piombo.

La simulazione di tali anticoincidenze avviene attraverso un calcolo
probabilistico della conversione dei gamma nel metallo.
Sono inoltre settabili differenti efficienze dell'apparato in funzione
di 13 differenti range di energia.

\subsection{Lo spettrometro magnetico}

\noindent
Quattro camere a fili costituiscono, insieme al magnete centrale, lo
spettrometro ma\-gne\-ti\-co.

La simulazione della risoluzione spaziale di tali camere e' ottenuta
attraverso uno smearing gaussiano con
\beq \sigma=103\s nm \eeq
delle coordinate trasverse effettive.

\subsection{L'odoscopio carico}

\noindent
L'odoscopio per la rivelazione dei decadimenti carichi, necessario per
una fase di pre-trigger, e' costituito da due semipiani verticali
adiacenti di materiale a scintillazione.
La simulazione avviene attraverso una semplice rivelazione del punto di
impatto del $\pi^\pm$ sul corrispondente semipiano.

\subsection{Il calorimetro adronico}

\noindent
Il calorimetro adronico rivela muoni $\mu^\pm$, pioni carichi $\pi^\pm$
e protoni $p$.

La simulazione di tale calorimetro fornisce l'energia del pione o del
protone con un errore di tipo gaussiano dato da:
\beq \frac{\sigma_E}{E}=\frac{d}{\sqrt{E}} \eeq
dove
\beq d=0.065 \eeq

La risoluzione in posizione e' data da uno smearing gaussiano della
posizione reale, usando una larghezza funzione dell'energia depositata:
\beq \sigma_{x,y}=\frac{c}{\sqrt{E\s[GeV]}} \eeq
dove
\beq c=15\s cm \eeq

\subsection{Il calorimetro elettromagnetico}

\noindent
Il calorimetro elettromagnetico rivela gamma, elettroni e pioni, frutto
del decadimento di \ks e \kl.

La simulazione del calorimetro fornisce l'energia depositata sulle
singole celle in accordo con la risoluzione del rivelatore:
\beq (\frac{\sigma_E}{E})^2=(\frac{a}{\sqrt{E}})^2+b^2 \eeq
dove
\beq a=0.035 \eeq
\beq b=0.005 \eeq

La risoluzione in posizione e' simulata da uno smearing gaussiano della
posizione reale, usando una larghezza funzione dell'energia depositata:
\beq \sigma_x=\frac{c_x}{\sqrt{E\s[GeV]}}\pm0.06\s cm \eeq
\beq c_x=0.42\s cm \eeq
\beq \sigma_y=\frac{c_y}{\sqrt{E\s[GeV]}}\pm0.06\s cm \eeq
\beq c_x=0.43\s cm \eeq

\subsection{L'odoscopio neutro}

\noindent
L'odoscopio neutro e' diviso, sia sul piano verticale, sia sul piano
orizzontale, in 64 strisce di materiale scintillatore.
La rivelazione dell'evento nell'odoscopio e' necessaria a dare lo start
alla catena di Trigger neutro.

\subsection{Il contatore di veto per i muoni}

\noindent
Il contatore di veto per i muoni consiste di tre piani di strisce di
scintillatore.

La simulazione di tale contatore tiene conto sia di un eventuale muone
rivelato, sia della sua posizione.

\section{Sciami elettromagnetici}
\setcounter{fig}{0}
\setcounter{tab}{0}

\noindent
Attraverso il programma NASIM-GEANT del CERN, vengono simulati
accuratamente gli sciami elettromagnetici prodotti dall'interazione dei
gamma con la materia.

La probabilita' di produzione di coppie $e^+ e^-$ da parte di un gamma
e' data da:
\beq P(x)=1-e^{-(\rho_i\sigma x)} \eeq
dove $x$ e' la distanza percorsa nel mezzo, $\rho_i$ la densita' di
centri di interazione del materiale, $\sigma$ la sezione d'urto per la
produzione di coppie.

Per quanto riguarda il calorimetro elettromagnetico, attraverso il
suddetto programma di simulazione, vengono create delle banche dati di
sciami elettromagnetici sviluppati nel Krypton liquido, in presenza di
criostato, elettrodi di rame disposti a zig-zag, separatori e
distanziatori collocati tra i vari elettrodi.
In corrispondenza di diverse energie e punti di impatto dei gamma,
esistono numerosi tipi di sciami, di modo da poter  simulare
realisticamente la deposizione di energia sul calorimetro
elettromagnetico.

Per una rapida analisi dei dati prodotti con metodo Montecarlo, e'stato
necessario ridurre la dimensione delle singole shower-box ad
$11\times11$ celle, inserendo un fattore di correzione per
tenere conto dell'energia persa con questo sistema di approssiamzione.
Il fattore di correzione stimato , ottenuto dal raffronto dei dati
generati e approssimati, risulta essere pari a:
\beq \alpha=1.091 \eeq
I risultati, uniti ad una notevole velocizzazione del programma, hanno
dato ragione a tale scelta.

\section{Trigger}
\setcounter{fig}{0}
\setcounter{tab}{0}

\noindent
Sempre all'interno del programma Montecarlo NMC sono presenti le
simulazioni di alcuni dei livelli di trigger previsti dall'esperimento
NA48:
\begin{itemize}
\item{\bf Trigger di livello 1} per un controllo rapido di
 anticoincidenze e contatori di veto disposti lungo la beam-line, e per
 una soppressione degli eventi con energia totale ricostruita inferiore
 a $20\s GeV$.
 Solo in seguito al consenso di qusto livello di trigger entrano in
 funzione i trigger di livello superiore; in questo modo si procede con
 una successiva riduzione del numero di eventi da analizzare, e con la
 possibilta' quindi di una ricostruzione piu' accurata del decadimento.
\item{\bf Trigger carico di livello 2} per la ricostruzione di energia,
 vertice di decadimento, massa invariante di \dpic, coordinate nelle
 camere a fili per i decadimenti carichi.
\end{itemize}

Presente all'interno di NMC, ma in versione alquanto semplificata,
e' il {\bf Trigger neutro di livello 2}, necessario alla ricostruzione
di energia, baricentro dell'energia, vertice di decadimento e numero di
gamma incidenti sul calorimetro elettromagnetico per i decadimenti
neutri.

Totalmente assente e' invece il {\bf Trigger di livello 2B}, \cite{L2B}
necessario alla distinzione, tra gli eventi a 5 e 6 gamma, dei decadmenti
in \tpiz da quelli in \dpiz con uno o due gamma accidentali.

  \chapter{Decadimenti neutri di $\bf K_S$ e $\bf K_L$}
\setcounter{fig}{0}
\setcounter{tab}{0}

\noindent
Una accurata analisi di decadimenti \ksdpiz, \kldpiz, \kltpiz prodotti
con metodo Montecarlo ha consentito di ottenere informazioni
dettagliate sulle varie quantita' misurabili per i tre modi di
decadimento.\\
Tali dati, cui e' stato necessario apportare alcune modifiche, sono
stati poi elaborati dal simulatore del Trigger neutro, al fine di
poter verificare, attraverso il raffronto delle quantita' ricostruite
con quelle generate, l'efficienza dell'apparato stesso.

\section{Analisi relativistica del decadimento}
\setcounter{fig}{0}
\setcounter{tab}{0}

\noindent
Dalla sola conoscenza dell'energia rilasciata dai gamma sulle singole
celle del calorimetro elettromagnetico, e' possibile ricostruire dati
utili al riconoscimento dell'evento, quali massa invarante delle
particelle in gioco, traiettorie e distribuzioni di energia dei pioni
e dei gamma originati dal decadimento di un \k.

\subsection{Massa invariante}

Dalla ricostruzione della massa invariante dei \piz e del \k, e' ad
esempio possibile distingure tra decadimenti \ksdpiz, \kldpiz, e
decadimenti invece \kltpiz con soli 4 gamma sul calorimetro: in
quest'ultimo caso,  quand'anche fosse possibile ricostruire esattamente
la massa invariante di due dei tre \piz, sarebbe invece impossibile
ricostruire la massa invariante del \k originario.
Allo stesso modo, dalla ricostruzione della massa invariante dei \piz e
del \k, e' possibile distinguere tra un decadimento di tipo \ksdpiz o
\kldpiz con 4 gamma sul calorimetro, e un decadimento neutro di
qualunque tipo con soli tre gamma sul calorimetro con la sovrapposizione
temporale di un gamma accidentale sincrono con gli altri tre.

Per ricostruire la massa invariante dei \piz e del \k dall'energia
depositata cella per cella, occorre innanzitutto riuscire a distinguere
il punto d'impatto sul calorimetro e l'energia dei singoli gamma.
Per questo occorre ricostruire esattamente le shower-box di ogni
gamma.\\
Tale ricostruzione risulta facile nel caso in cui le varie shower-box
non siano spazialmente sovrapposte, e viene eseguita attraverso la
ricerca del massimo di energia depositata sulle singole celle e il
calcolo della derivata dell'energia rispetto alla posizione, al fine di
trovare i confini della singola shower-box.
La distribuzione dell'energia in funzione della posizione trasversa alla
beam-line dello sciame elettromagnetico e' infatti nota
sperimentalmente, e risulta dalla somma di tre esponenziali decrescenti,
rappresentanti la testa, il corpo, e la coda dello sciame, con simmetria
circolare.\\
Ponendo un limite inferiore alla derivata di tale funzione, e' possibile
individuare i confini della singola shower-box senza il rischio di
conteggiare anche celle in cui la deposizione di energia sia puramente
frutto di un basso noise interno o esterno all'apparato di rivelazione.

Una volta ricostruite con precisione energia e punto di impatto dei
gamma, si procede con la ricerca delle coppie di gamma che, in quanto a
massa invariante, possono provenire dal decadimento dello stesso \piz.
Definendo un intervallo di accettazione per la massa invariante del \piz
in funzione dell'efficienza e della precisione del rivelatore e del
livello a cui si ricostruisce la quantita' fisica in esame, la massa del
primo \piz e la massa del secondo \piz, nel caso di decadimenti \ksdpiz
e \kldpiz, risultano essere correlate negativamente, con un diagramma di
dispersione degli eventi di forma ellissoidale, con asse maggiore,
inclinato di $-45^\circ$ rispetto alla verticale, direttamente
proporzionale ai limiti imposti sulla massa del singolo \piz, e asse
minore direttamente proporzionale alla risoluzione sulla misura
dell'energia da parte dell'apparato di rivelazione.\\
Per una buona ricostruzione degli eventi, i limiti inferiore e superiore
alla massa del \piz ($0.135\s GeV$) sono fissati in $0.134\s GeV$ e
$0.136\s GeV$ rispettivamente.
Da una analisi dei dati ottenuti, il diagramma di dispersione, per i
soli eventi a 4 gamma sul calorimetro, risulta avere una larghezza di
$383\s KeV$, corrispondente a $3.8\cdot10^{-3}\s m(\pi^0)$, dello
stesso ordine di grandezza della risoluzione in energia del calorimetro
eletromagnetico
\beq \frac{\sigma_E}{E}=\frac{0.035}{\sqrt{E}} \eeq

\subsection{Distribuzione di gamma e pioni}

In seguito alla identificazione delle singole particelle (pioni e gamma)
prodotte dal decadimento del \k, si riesce a ricostruire la traiettoria
di volo virtuale dei \piz (virtuale perche' il \piz decade subito, in
conseguenza di una vita media estremamante breve) e l'angolo di apertura
delle coppie di gamma provenienti dal singolo \piz.
La traiettoria di volo del \piz viene costruita come congiungente del
vertice di decadimento del \k con il baricentro dell'energia dei due
gamma provenienti dal \piz.\\
Per quanto riguarda l'angolo di apertura dei gamma, questo e' stato
calcolato nel riferimento del laboratorio, e ricostruito nel riferimento
della particella \piz.

Nel riferimento del laboratorio sono state ottenute le seguenti
distribuzioni, rispettivamente per l'angolo di apertura di ciascuna
coppia di gamma, e per l'angolo rispetto alla beam-line della
traiettoria del \piz:\\

\bce
\addtocounter{fig}{1}
Fig.\fig: \bksdpiz Angolo di apertura di gamma e \piz
\ece
\vsp{0.5cm}

\bce
\addtocounter{fig}{1}
Fig.\fig: \bkldpiz Angolo di apertura di gamma e \piz
\ece
\vsp{0.5cm}

\bce
\addtocounter{fig}{1}
Fig.\fig: \bkltpiz Angolo di apertura di gamma e \piz
\ece

Per il decadimento \kltpiz a tre corpi, l'angolo medio di apertura dei
gamma risulta maggiore del corrispondente angolo per il decadimento
\kldpiz a due corpi; i gamma, 6 invece di 4, sono mediamente meno
energetici, e per la conservazione del quadriimpulso totale del sistema,
l'angolo $\theta_{ij}$ di apertura dei due gamma deve essere mediamente
maggiore.
La conservazione del quadriimpulso totale implica
\beq {m_{\pi^0}}^2=\sum E_{\gamma_i} E_{\gamma_j} (1-\cos\theta_{ij})
\eeq

Sempre per il decadimento \kltpiz, l'angolo tra beam-line e traiettoria
virtuale del \piz risulta minore del corrispondente angolo per il
decadimento \kldpiz; per la conservazione del quadriimpulso totale del
sistema ($n$ e' il numero di \piz):
\beq {m_{K}}^2=\sum E_{\pi^0_i} E_{\pi^0_j} (1-\cos\varphi_{ij})
+n{m_{\pi^0}}^2 \eeq
dove $\varphi_{ij}$ e' mediamente doppio dell'angolo in questione,
e tenendo conto che i 3\piz, seppur singolarmente meno energetici, sono
in numero maggiore rispetto ai \piz di un decadimento \kldpiz, l'angolo
$\varphi_{ij}$ deve essre mediamente inferiore.

Nel riferimento della particella \piz, l'angolo totale di apertura di
ciascuna coppia di gamma, ottenuto attraverso una trasformazione di
Lorentz delle coordinate e degli impulsi del laboratorio, risulta pari
a $180^\circ\pm2.7^\circ$, in accordo con un decadimento a due corpi.

\clearpage

Punto di impatto ed energia dei singoli gamma sul calorimetro risultano
cosi' distribuiti:\\

\bce
\addtocounter{fig}{1}
Fig.\fig: \bksdpiz Punto di impatto e Energia gamma
\ece
\vsp{0.5cm}

I grafici si riferiscono a 10000 eventi per ciascuno dei tre modi di
decadimento; per quanto riguarda i decadimenti \ksdpiz occorre tenere
presente che solo nel $62.4\%$ dei casi almeno un gamma riesce a
raggiungere il calorimetro elettromagnetico, a causa della presenza di
una anticoincidenza anti-\ks nell'apparato.

\bce
\addtocounter{fig}{1}
\nopagebreak
Fig.\fig: \bkldpiz Punto di impatto e Energia gamma
\ece
\vsp{0.5cm}

\bce
\addtocounter{fig}{1}
\nopagebreak
Fig.\fig: \bkltpiz Punto di impatto e Energia gamma
\ece
\vsp{0.5cm}

Da una analisi dell'energia dei singoli gamma in funzione del punto
d'impatto sul calorimetro, i gamma piu' energetici risultano fortemente
concentrati verso il centro del calorimetro, in accordo con la
conservazione del quadriimpulso totale del sistema ($E_\gamma$ maggiore
implica infatti $\theta_\gamma$ minore).
La distribuzione dell'energia del gamma in funzione della distanza dal
centro del calorimetro del punto d'impatto risulta essere:\\

\bce
\addtocounter{fig}{1}
Fig.\fig: \bksdpiz Energia gamma vs. distanza dal centro
\ece
\vsp{0.5cm}

\bce
\addtocounter{fig}{1}
Fig.\fig: \bkldpiz Energia gamma vs. distanza dal centro
\ece
\vsp{0.5cm}

\bce
\addtocounter{fig}{1}
Fig.\fig: \bkltpiz Energia gamma vs. distanza dal centro
\ece
\vsp{0.5cm}

\section{Analisi off-line}
\setcounter{fig}{0}
\setcounter{tab}{0}

\noindent
L'analisi dell'evento effettuata a partire dalla conoscenza dell'energia
depositata su ciascuna cella del calorimetro elettromagnetico potra'
essere in seguito utilizzata per una analisi off-line degli eventi
registrati su supporto magnetico, attraverso un sistema di compattazione
dei dati e di zero suppression \cite{data}.\\
Tale analisi consentira' di eliminare quasi totalmente gli eventi di
fondo per entrambi i modi di decadimento, carico e neutro, e di
ricostruire gli eventi utili alla misura del doppio rapporto $R$ con la
precisione richiesta.

  \chapter{Simulazione del Trigger neutro}

\section{Il rumore}
\setcounter{fig}{0}
\setcounter{tab}{0}

\noindent
Essenziale ai fini di una simulazione realistica degli eventi e' la
trattazione del rumore.\\
Internamente all'apparato esistono fonti di rumore identificabili
essenzialmente negli amplificatori, preamplificatori e shaper collegati
alle singole celle del calorimetro.
Questo viene simulato attraverso un noise con distribuzione gaussiana a
media $\mu=0$ e sigma $\sigma=30\s MeV$, corrispondente al noise
massimo sperimentalmente misurato su ogni singola cella del calorimetro;
la media $\mu=0$ e' frutto di un noise fluttuante nel tempo a media
nulla.\\
I noises relativi ai $128\times128$ canali del calorimetro vengono
variati ad ogni ciclo di clock rigenerando ogni volta $128\cdot128$
numeri distribuiti in modo gaussiano.

\section{Amplificatori}
\setcounter{fig}{0}
\setcounter{tab}{0}

\noindent
Il sistema di lettura del calorimetro elettromagnetico e' costituito da
due sezioni: la prima analogica, la seconda digitale.\\
Un amplificatore amplifica il segnale proveniente da ciascuna cella del
calorimetro; il segnale viene poi inviato ad uno shaper per formare
l'impulso da analizzare.

Il segnale di ciascuna cella, formato dalla raccolta di corrente sugli
elettrodi che fungono da condensatore, ed amplificato
dall'amplificatore, e' di forma triangolare, ed e' caratte\-riz\-zato da
una ampiezza di $2.4\mu A/GeV$, da un tempo di salita di $3\s \mu s$,
e da un tempo di discesa, corrispondente al tempo di scarica del
condensatore, di circa $50\s \mu s$ \cite{shaper}.

La simulazione degli amplificatori richiede una banca dati contenente i
fattori di amplificazione, con massimo normalizzato ad uno, relativi ad
ogni singolo canale, con la possibilita' di tenere conto di eventuali
malfunzionamenti, parziali o totali, di alcuni canali.
Non essendo ancora disponibile, in quanto non ancora effettuata, una
mappatura completa dei singoli canali, viene adottata una distribuzione
di tipo gaussiano dei $128\times128$ fattori di amplificazione, con
media $\mu=1$ e sigma $\sigma=0.1$.\\

\bce
\addtocounter{fig}{1}
Fig.\fig: Amplificazione
\ece
\vsp{0.5cm}
Tale $\sigma$ e' in accordo con l'errore introdotto dai condensatori
utilizzati negli amplificatori.

Ogni canale risulta poi essere sfasato, sempre a causa delle piccole
diversita' dei vari componenti elettronici, rispetto al canale
adiacente, con una distribuzione dei ritardi canale-canale di tipo
gaussiano, con media $\mu=0$ e sigma $\sigma=1\s ns$, corrispondente
allo sfasamento medio dei canali l'un l'altro.\\

\bce
\addtocounter{fig}{1}
Fig.\fig: Ritardi canale-canale
\ece
\vsp{0.5cm}

\section{Shaper}
\setcounter{fig}{0}
\setcounter{tab}{0}

\noindent
Funzione dello shaper e' quella di sagomare  il segnale per una
successiva analisi.
In corrispondenza ad una energia depositata sulla singola cella di
$10\s GeV$, il massimo dell'uscita dello shaper e' di circa $2\s V$
\cite{LKr}.
La forma d'onda del segnale in uscita dallo shaper, calcolata per via
teorica attraverso la funzione di trasferimento $in/out$ dell'apparato,
e ottenuta da fit di dati sperimentali, e' caratterizzata da una altezza
proporzionale alla derivata del segnale proveniente dall'amplificatore,
da una larghezza a meta' altezza pari a $60\s ns$, e da un undershoot
del $3\%$ di durata $3\s \mu s$ \cite{shaper}.\\

\bce
\addtocounter{fig}{1}
Fig.\fig: Shaper: Amp. vs. Time (ns)
\ece
\vsp{0.5cm}

Essendo lo shaper sensibile alla variazione del segnale in
ingresso (rivela infatti la derivata di tale segnale), risponde
immediatamente all'impulso amplificato proveniente dalla cella del
calorimetro, e in modo del tutto indipendente dalla durata del segnale
in ingresso.\\
Pertanto, in caso di sovrapposizione temporale di piu' eventi sulla
singola cella, lo shaper ad essa collegato riesce a distinguerli entro
l'intervallo di durata del segnale di shaper, ovvero $3\s \mu s$.
In realta', considerando il fatto che l'undershoot di tale segnale e'
relativamente basso ($3\%$), la sovrapposizione di un secondo segnale
con l'undershoot del primo non incide in maniera rilevante ai fini della
rivelazione dell'evento, almeno al fine di un conteggio dei gamma
incidenti sul calorimetro.
Tali sovrapposizioni avvengono comunque molto raramente, essendo il
rate previsto di eventi sulla singola cella inferiore ad $1\s KHz$.

La sezione di shaper richiede particolare cura e attenzione nella
simulazione.
La struttura delle $128\times128$ celle del calorimetro, prodotta sotto
forma di matrice, viene fornita di una terza dimensione, quella
temporale.
I metodi utilizzati sono diversi a seconda del tipo di analisi da
effettuare.

Un primo sistema, estremamente veloce ma leggermente riduttivo, consiste
in una a\-na\-li\-si temporalmente limitata a $200\s ns$ intorno al
picco in energia rivelato, e nella soppressione dell'undershoot, con una
conseguente concentrazione degli eventi, compatibilmente con i vari
stadi di pipeline dell'elettronica, ed una riduzione dei tempi necessari
all'analisi.\\
Un secondo sistema, completo ed esatto, consiste invece nell'analisi
del segnale dello shaper per tutta la sua durata temporale, tenendo
cosi' conto anche di eventuali sovrapposizioni di due o piu' segnali.

Nonostante questa sezione lavori ancora in modo continuo, e' possibile
simularla campionando il segnale in funzione di una successiva lettura
da parte di una sezione digitale regolata da clock.\\
La durata del campionamento, nella simulazione dell'apparato, e' quindi
funzione del tipo di analisi da effettuare: 8 campionamenti sono
sufficienti a coprire temporalmente il picco del segnale, mentre ne
occorrono circa 125 per coprire anche tutto il periodo corrispondente
all'undershoot, con un rallentamento notevole del programma.

La forma d'onda del segnale di shaper, nota analiticamente, viene
inizialmente cam\-pio\-na\-ta a step di $1\s ns$ e registrata in una
banca dati, cosi' da rendere piu' veloce il sistema di simulazione.
Il calcolo attraverso la funzione di trasferimento richiede infatti
tempi di esecuzione quattro volte superiori.

Ad ogni step, corrispondente ad un tempo di $25\s ns$, viene letto il
valore dell'energia depositata in ogni cella, convertito in Volt,
moltiplicato per il fattore di amplificazione e sfasato del ritardo
relativo al corrispondente canale, moltiplicato per la fase dovuta alla
lettura da parte dell'elettronica, e sporcato dal rumore relativo al
canale, quindi viene creato il vettore contenente la campionatura del
segnale come formato dallo shaper, nel caso di energia diversa da zero,
e shiftato di una posizione temporale.\\
Nel caso di sovrapposizione temporale di ue o piu' segnali sulla stessa
cella, alla nuova campionatura viene aggiunto, step per step, il
corrispondente valore precedentemente re\-gi\-stra\-to.
Per le celle ove non sia depositata energia il processo e' notevolmente
semplificato, e consiste in un semplice shift dei vari piani
temporali della matrice di shaper, cosi' da poter simulare correttamente
i vari stadi di pipeline.

Un differente sistema di simulazione, quale ad esempio una campionatura
successiva del segnale, avrebbe richiesto tempi di esecuzione maggiore,
e non avrebbe comunque apportato nessuna variazione agli stadi
successivi.

\section{Il clock}
\setcounter{fig}{0}
\setcounter{tab}{0}

\noindent
La sezione digitale dell'apparato di trigger, strutturata in pipeline,
e' progettata per funzionare con un periodo di clock di $25\s ns$, pari
ad una frequenza di $40\s MHz$.
Tale clock, totalmente indipendente dall'arrivo dell'evento sul
calorimetro, non consente una cam\-pio\-na\-tu\-ra del segnale di shaper
sincrona col tempo relativo al proprio massimo, ma fornisce una
campionatura con fase distribuita in maniera casuale uniforme in un
intervallo di $25\s ns$ rispetto al tempo del picco.\\
In generale non e' quindi disponibile la lettura del massimo del
segnale, ma un valore compreso tra  il $75\%$ e il $100\%$ del massimo.
Compito di una parte dell'apparato sara' poi quello di ricostruire,
attraverso una interpolazione di tipo parabolico, il valore del picco,
corrispondente al valore dell'energia rilasciata sul calorimetro.

\section{Struttura pipelined}
\setcounter{fig}{0}
\setcounter{tab}{0}

\noindent
Il rate elevato di eventi sul calorimetro richiede un sistema di lettura
veloce per l'apparato di rivelazione.\\
La realizzazione di un trigger tradizionale, che blocca cioe'
l'acquisizione di qualsiasi evento per tutto il tempo necessario
all'analisi dell'ultimo evento acquisito, avrebbe posto seri limiti alla
velocita' di lettura e all'efficienza dell'apparato stesso.
L'utilizzo di una struttura pipelined consente di ovviare a questo
problema.

La pipeline consiste nella suddivisione di una singola rete combinatoria
in piu' blocchi, attraverso l'interposizione di registri \cite{F.Rossi};
ciascun blocco realizza una parte dell'operazione cui l'intera rete e'
preposta, permettendo un'acquisizione piu' veloce dei dati, e quindi
una minimizzazione dei tempi morti.

E' possibile osservare che, dopo un periodo iniziale pari a $n$ periodi
di clock, se $n$ e' il numero di stadi di pipeline, i risultati
dell'operazione compiuta dal circuito vengono forniti con frequenza
pari a quella del clock.

La simulazione dei vari stadi di pipeline, ridotti al numero minimo
necessario per una analisi dell'evento da parte del programma, avviene
attraverso uno shift temporale delle varie quantita' memorizzate, ovvero
attraverso uno shift degli indici di ciascuna variabile.

\section{I sommatori analogici}
\setcounter{fig}{0}
\setcounter{tab}{0}

\noindent
I segnali provenienti dagli amplificatori e dagli shaper connessi ad
ogni singola cella del calorimetro elettromagnetico vengono sommati a
gruppi di $8\times2$ in verticale e in orizzontale, ottenendo cosi' una
doppia mappatura (nella vista $x$ e nella vista $y$) concentrata del
calorimetro.

\section{I digitalizzatori}
\setcounter{fig}{0}
\setcounter{tab}{0}

\noindent
Il segnale ottenuto dai sommatori analogici viene passato alla sezione
digitale del Trigger neutro.
Innanzitutto il segnale viene digitalizzato attraverso Fast ADC a 10 bit
che lavorano ad una frequenza di $40\s MHz$.
In uscita agli ADC il valore corrispondente all'energia rilasciata sul
singolo blocchetto $8\times2$ celle si presenta come numero di step di
digitalizzazione, piu' un livello di zero settato a 100 step di ADC.

L'algoritmo impiegato e':
\beq step=\frac{(max-min)}{(2^{nbit}-levZero-1)} \eeq
\beq Digit=\frac{Analog-mod(Analog,step)}{step}+levZero  \eeq
dove $max$ corrisponde al livello massimo di saturazione del singolo
Fast ADC ($100\s GeV$), $min$ e' fissato a $0$, $step$ corrisponde
all'ampiezza del singolo passo di digitalizzazione, $nbit$ e' il numero
di bit dei Fast ADC, $levZero$ e' il livello di piedistallo.

Il piedistallo di 100 step di digitalizzazione e' introdotto negli ADC
per poter tenere conto dell'undershooot dei segnali in alcuni tipi di
analisi; tale livello viene invece soppresso attraverso un semplice
sottrattore di piedistallo (baseline restoring)
\beq Digit=Digit-levZero \eeq
\beq if \s (Digit.lt.levZero) \s\s Digit=0 \eeq
per ripristinare lo zero del segnale per la successiva analisi da parte
del Trigger neutro.

Le variabili sono trattate come Real per sfruttare al meglio le
potenzialita' dei compilatori Fortran, quindi velocizzare il processo,
senza perdere in precisione: l'unica differenza infatti nella
registrazione in memoria di un numero Real rispetto ad un Integer sta
nel fatto che il primo e' caratterizzato da virgola mobile, conservando
pero' un numero maggiore di cifre significative nella propria mantissa.

Il procedimento di digitalizzazione, come pure la somma analogica in
blocchetti $8\times2$ , avviene separatamente per le due viste $x$ e
$y$.

Considerando un segnale in ingresso distribuito in modo casuale uniforme
tra $0$ e $100\s GeV$, ed un numero di bit dell'ADC pari a 10, il
rapporto tra numero digitalizzato e analogico
\beq \frac{Digit}{Analog} \eeq
risulta avere una ditribuzione siffatta:

\bce
\addtocounter{fig}{1}
Fig.\fig: Quantizzazione ADC
\ece
\vsp{0.5cm}
con un evidente effetto di quantizzazione, presente in maniera
predominante sui valori inferiori.\\
E' lecito quindi aspettarsi una perdita in precisione sulle basse
energie, poco rilevante comunque per l'identificazione dell'evento e per
la successiva analisi effettuata dal trigger.

\section{I filtri}
\setcounter{fig}{0}
\setcounter{tab}{0}

\noindent
Il segnale digitalizzato viene inviato ad un sistema di analisi,
costituito dal filtro temporale e dalla maschera spaziale.\\
Entrambi questi sistemi lavorano, indipendentemente sulle due
proiezioni, sulle somme digitali dei singoli canali (d'ora in poi sara'
utilizzata la notazione canale anziche' blocchetto di $8\times2$ celle).

Il filtro temporale abilita il singolo canale, quello temporalmente
successivo e i due precedenti, nel caso in cui il valore digitalizzato
risulti superiore ad una soglia settata intorno a $0.5\s GeV$, ovvero
poco al di sopra del valore massimo di noise raggingibile sul singolo
canale, e corrispondente a circa $30\s MeV\cdot16=0.48\s GeV$.
Questo comporta una riduzione del rumore, abilitando soltanto i gruppi
di celle che possono effettivamente contenere un picco in energia.

Una ulteriore riduzione del rumore viene operata attraverso una
selezione dei canali che devono contribuire alla successiva somma in
righe e colonne del calorimetro elettromagnetico.
Questa selezione viene eseguita dalla maschera spaziale, che,
confrontando il segnale proveniente da ciascun canale con una soglia
prestabilita di $0.5\s GeV$, abilita alla somma esclusivamente quei
canali per cui il segnale risulta essere sopra soglia, e un certo numero
di canali adiacenti, a seconda della configurazione selezionata.\\
L'utilizzo della maschera spaziale nell'esperimento NA48 e' tuttora
incerto.

La simulazione di entrambi questi apparati, non complessa dal punto di
vista concettuale, richiede particolare attenzione nella ricostruzione
dei vari stadi di pipeline, con un continuo aggiornamento dei canali
abilitati alle somme in righe e colonne.

\section{I sommatori in righe e colonne}
\setcounter{fig}{0}
\setcounter{tab}{0}

\noindent
I canali abilitati dal filtro temporale e dalla maschera spaziale
vengono sommati in righe e colonne separatamente nelle due viste $x$
e $y$, ottenendo in questo modo 64 valori corrispondenti alle 64
proiezioni per ciascuna vista.

I sommatori digitali a 12 bit eseguono ad ogni step temporale di $25\s
ns$ la somma dei vari canali, tenendo conto delle rispettive
abilitazioni per i vari stadi di pipeline;
quattro sono infatti le informazioni temporali necessarie al Peak-Finder
per la ricostruzione di un picco.\\
La variazione dell'abilitazione fornita dai filtri in slot temporali
successive e precedenti quella in esame, obbliga ad eseguire la somma
ogni volta per vari stadi di pipeline.

Ricostruendo l'evoluzione temporale di una delle 64 colonne, tenuto
conto di sfasamento e fattore di amplificazione delle singole celle,
fase del clock rispetto al tempo del picco, filtro temporale, e
raffrontando il segnale cosi' ottenuto col segnale di uno degli shaper,
si ottiene:

\bce
\addtocounter{fig}{1}
Fig.\fig: Colonna digitalizzata e Shaper
\ece
\vsp{0.5cm}

L'informazione originaria sull'energia depositata su ogni singola cella
e' adesso compattata in informazioni sull'energia depositata su una
colonna o riga di $128\times2$ celle, con una inevitabile perdita in
risoluzione, ma con una conseguente notevole velocizzazione del sistema
di trigger, cosi' da rendere possibile una veloce analisi dei dati,
compatibile con il rate di eventi sul calorimetro elettromagnetico.

Una ricostruzione precisa dell'evento, con l'acquisizione dei dati
relativi alle singole celle, viene effettuata alla fine della catena di
trigger, nel caso in cui quest'ultimo riveli l'evento come candidato ad
essere un decadimento \ksdpiz, \kldpiz, e dia quindi l'abilitazione alla
registrazione dell'evento per una successiva analisi off-line.

\section{Il Peak-Sum}
\setcounter{fig}{0}
\setcounter{tab}{0}

\noindent
Elemento di fondamentale importanza nell'apparato di Trigger neutro e'
il Peak-Sum \cite{F.Rossi}, progettato per la ricostruzione,
separatamente nelle due viste, di energia e momenti di ordine 1 e 2
dell'energia rispetto alla posizione proiettata.

Il Peak-Sum elabora gli ingressi costituiti dalle 64 somme in righe e
colonne attraverso l'impiego di singoli chips, progettati per poter
lavorare in differenti modi, ciascuno dei quali vede quattro canali.
Sedici chips sono quindi necessari per l'analisi di 64 canali.

Gli algoritmi impiegati sono:
per la ricostruzione dell'energia (momento di ordine 0):
\beq M_{0X}=\sum E_{i,x} \eeq
\beq M_{0Y}=\sum E_{i,y} \eeq
per la ricostruzione dei momenti di ordine 1:
\beq M_{1X}=\frac{\sum E_i X_i}{\sum E_i} \eeq
\beq M_{1Y}=\frac{\sum E_i Y_i}{\sum E_i} \eeq
per la ricostruzione dei momenti di ordine 2:
\beq M_{2X}=\frac{\sum E_i X_i^2}{\sum E_i} \eeq
\beq M_{2Y}=\frac{\sum E_i Y_i^2}{\sum E_i} \eeq
dove $X_i$ e $Y_i$ sono rispettivamente il numero di colonna e riga
nelle proiezioni $x$ e $y$, ed $E_i$ l'energia ricostruita sul
rispettivo canale.

Il momento di ordine 0 corrisponde esattamente all'energia depositata e
ricostruita sulla singola vista.\\
Il momento di ordine 1 corrisponde al baricentro dell'energia depositata
sul calorimetro elettromagnetico.\\
Il momento di ordine 2 viene utilizzato, unitamente ai precedenti, per
il calcolo del vertice di decadimento della particella, attraverso la
formula
\beq Z_V=D-\frac{E}{m_K}\sqrt{(M_{2X}+M_{2Y})-(M_{1X}^2+M_{1Y}^2)} \eeq
dove $Z_V$ e' la distanza del vertice di decadimento dalla targhetta
\ks, $D$ la distanza di tale targhetta dal calorimetro elettromagnetico
($122.35\s m$), $E$ l'energia totale ricostruita, $m_k$ la massa del \k
($0.497\s GeV$).
La formula per il calcolo del vertice di decadimento $Z_V$ e' data dalla
conservazione del quadriimpulso totale del sistema, nell'approssimazione
di piccoli angoli , e dalla conoscenza della massa invariante del \k.

Energia, baricentro dell'energia, vertice di decadimento della
particella sono, insieme all'informazione sul numero di picchi rivelati
sul calorimetro elettromagnetico, le quantita' fisiche dall'analisi
delle quali e' possibile ricostruire il tipo di decadimento, e rimuovere
quindi il fondo dovuto ad eventi accidentali e ai decadimenti \kltpiz.

Le somma per i momenti di ordine 0,1,2 dell'energia rispetto alla
posizione proiettata vengono effettuate da sommatori floating point
\cite{chip} a 9+1 bit di mantissa, ed esponente di 3,4,5 bit
rispettivamente per i tre momenti, per garantire una non saturazione del
risultato.
La registrazione dei soli 9 bit successivi al primo 1 nel numero in
formato binario consente di recuperare una cifra significativa del
numero binario in questione, lasciando pressoche' inalterati i circuiti
sommatori.

Il procedimento con cui vengono effettuate le somme all'interno del
Peak-Sum si avvale di chip appositamente adibiti alla ricombinazione dei
risultati provenienti dalle otto schede per ciascuna vista che
compongono l'apparato, ciascuna delle quali comprensiva di due chips per
il trattamento di quattro canali ciascuno.

La simulazione di tale struttura, lavorando i computer su numeri a 32 o
64 bit, richiede una elaborata conversione del numero digitale a 12 bit
in un numero floating point con mantissa di 10 bit, ed esponente di
3,4,5 bit a seconda dell'ordine del momento ricostruito.
Per questa conversione sono create due apposite procedure, impiegabili a
seconda del tipo di calcolatore e di compilatore utilizzati.

Un primo e piu' semplice sistema consiste nel calcolo del numero di bit
necessari in formato binario a contenere il numero in esame, attraverso
il successivo confronto col valore $2^n-1$, corrispondente al massimo
numero registrabile in $n$ bit.
l numero digitale viene quindi arrotondato con un numero di cifre
significative pari a 10 attraverso l'algoritmo:
\beq floating \s point=digit-mod\s(digit,\s2^{\s n-10}) \eeq
nel caso in cui $n$ sia maggiore di 10, e lasciato inalterato nel caso
in cui $n$ sia minore di 10.
I numeri cosi' trattati vengono nuovamente registrati in memoria come
Real.

Il secondo metodo per la conversione di un numero qualsiasi in floating
point consiste nella costruzione di apposite maschere, di 26 o 58 bit a
seconda del calcolatore e del compilatore utilizzati, dove un solo bit
in posizione $n$ e' uguale a 1, ed i restanti sono nulli.\\
Il numero digitale, registrato sotto forma di Real a 32 (64) bit, viene
confrontato con le 26 (58) maschere per identificare, attraverso
operazioni logiche di tipo $AND$ e $OR$, la posizione del primo bit
diverso da zero nella mantissa, e quindi ricostruire la mantissa del
numero in esame.
L'esponente e' invece fornito direttamente dal calcolo del logaritmo in
base 2 del numero.\\
I calcolatori lavorano internamente in base 16, e l'incremento
(diminuzione) di una unita' dell'esponente corrisponde quindi ad uno
shift verso sinistra (destra) della virgola di quattro posizioni;
non potendo eccedere il numero di bit previsti, ne' perdere in cifre
significative, il compilatore procede ad una riallocazione dei bit
(normalizzazione), in modo tale da riportare il primo 1 in una delle
prime quattro posizioni, e all'incremento o di\-mi\-nu\-zio\-ne di una
unita' dell'esponente.
Considerando che i compilatori Fortran registrano i numeri in formato
Real in modo tale da disporre il primo 1 in uno dei primi quattro bit,
sono sufficienti solo quattro raffronti per ricostruire esattamente la
mantissa.\\
Questo procedimento consente di estrarre dal numero Real nella memoria
del calcolatore i primi 9 bit successivi al primo bit diverso da zero.

La conversione da Real (26 bit di mantissa + 6 bit di esponente) a
floating point (10 bit di mantissa + 3,4,5 bit di esponente) viene
effettuata dopo ogni somma in uscita da ciascuno dei 16 chips,
indipendentemente per i momenti dell'energia di ordine 0,1,2.

\section{Il Peak-Finder}
\setcounter{fig}{0}
\setcounter{tab}{0}

\noindent
Lo stesso chip \cite{F.Rossi}che provvede alla realizzazione del
Peak-Sum puo' essere programmato per la ricerca dei picchi di energia
rilasciata sul calorimetro elettromagnetico, al fine di ricostruire il
numero di gamma incidenti sul rivelatore.

Perche' un picco venga rivelato come tale occorre che questo risulti
dalle analisi combinate spaziale e temporale.

L'algoritmo di ricerca spaziale sui 64 canali rivela un picco in
$X_i(0)$ nel caso in cui:
\beq X_i(0)\geq X_{i+1}(0) \eeq
\beq X_i(0) >   X_{i-1}(0) \eeq
Il valore in basso a destra si riferisce al numero del canale (riga o
colonna), mentre la quantita' tra parentesi rappresenta il tempo
dell'evento campionato ogni $25\s ns$.

L'algoritmo di ricerca temporale sui 64 canali rivela un picco in
$X_i(0)$ se:
\beq X_i(0)\geq X_i(+1) \eeq
\beq X_i(0) >   X_i(-1) > X_i(-2) \eeq

Altra condizione imposta per l'identificazione di un picco su un canale
e' che nella slot temporale precedente non fosse presente alcun picco
nei due canali adiacenti.
Questa condizione e' dovuta al fatto che una serie di gamma
temporalmente separati di $25\s ns$, e spazialmente contigui di cella,
paralizzerebbero l'apparato di trigger, non consentendo quindi l'analisi
di nessuno di questi eventi; in un caso del genere infatti esisterebbe
l'abilitazione dell'algoritmo di ricerca temporale, ma verrebbe a
mancare con buona probabilita' l'abilitazione da parte dell'algoritmo di
ricerca spaziale.

La simulazione software del sistema Peak-Finder si avvale, per la
ricerca dei picchi, dei dati registrati nei successivi stadi di
pipeline, limitabili ad un numero minimo di cinque per una accurata
simulazione dell'apparato.

Le informazioni sul numero di picchi e sulla loro posizione, vengono
registrate in apposite banche dati ad accesso rapido.

\section{La temporizzazione dell'evento}
\setcounter{fig}{0}
\setcounter{tab}{0}

\noindent
Lo stesso chip impiegato per la ricostruzione dei momenti di ordine
0,1,2 dell'energia rispetto alla posizione proiettata e per la ricerca
dei picchi e' programmabile per la ricostruzione del tempo del picco in
sottoslot di $3.125\s ns$, intervallo corrispondente a $1/8$ del periodo
di clock.

Attraverso una interpolazione rettilinea dei valori campionati ogni
$25\s ns$, l'analisi procede con una ricerca della sottoslot temporale
in cui cade il tempo di riferimento.
Tale tempo di riferimento corrisponde, a seconda della scelta
programmabile via VME, al tempo in cui il segnale raggiunge
rispettivamente $\frac{1}{2}$, $\frac{1}{2}+\frac{1}{16}$,
$\frac{1}{2}+\frac{1}{8}$, $\frac{1}{2}+\frac{1}{8}+\frac{1}{16}$
dell'altezza del massimo campionato, per una migliore precisione nella
temporizzazione del segnale.

In funzione della differenza tra i tempi relativi ai vari picchi
rivelati, il Trigger Supervisor dara' poi l'abilitazione
all'acquisizione dell'evento.

Da una analisi effettuata su 30000 eventi con gamma sincroni sul
calorimetro elettromagnetico, emerge la seguente distribuzione di
distanza temporale massima tra i tempi relativi ai vari picchi, espressa
in numero di sottoslot da $3.125\s ns$ ciascuna:

\bce
\addtocounter{fig}{1}
Fig.\fig: Distribuzione $\Delta\s T$ dei gamma
\ece
\vsp{0.5cm}

Tali risultati suggeriscono quindi un intervallo temporale di
accettazione per eventi buoni corrispondente a $(0,3)$ sottoslot di
differenza massima tra i tempi relativi ai vari gamma.

\section{Look-up table}
\setcounter{fig}{0}
\setcounter{tab}{0}

\noindent
I moduli Look-up table sono costituiti essenzialmente da memorie
programmabili, e permettono l'elaborazione dei risultati forniti dal
Peak-Sum e dal Peak-Finder per il calcolo delle quantita' richieste dal
sistema di trigger \cite{LUT}.

Le funzioni realizzate da questi moduli riguardano l'eleborazione dei
dati relativi alle proiezioni $x$ e $y$, e consentono di determinare:
\begin{itemize}
\item tempo esatto dell'evento sul calorimetro elettromagnetico
\item energia totale depositata sul calorimetro elettromagnetico
\item posizione del baricentro dell'energia sul calorimetro
\item posizione del vertice di decadimento della particella
\item numero di picchi sul calorimetro elettromagnetico
\end{itemize}

La ricostruzione del valore dell'energia depositata sul calorimetro
elettromagnetico avviene attraverso una interpolazione parabolica del
valore del massimo e dei due va\-lo\-ri temporalmente contigui.

\section{Il Trigger Supervisor}
\setcounter{fig}{0}
\setcounter{tab}{0}

\noindent
Una serie di tagli sulle quantita' calcolate viene effettuata dal
Trigger Supervisor, in modo da ridurre il rate di eventi candidati ad
essere registrati su supporto magnetico.

I dati generati da NMC e quelli ricostruiti dal trigger relativi agli
eventi analizzati vengono registrati su disco, per poter poi calcolare
l'efficienza dell'apparato sperimentale in corrispondenza di diversi
tagli effettuati.

  \chapter{Test e risultati}

\section{Campionamento del segnale ed errori di arrotondamento}
\setcounter{fig}{0}
\setcounter{tab}{0}

\noindent
Attraverso la simulazione dettagliata di tutto l'apparato sperimentale,
costituito da ri\-ve\-la\-tore piu' trigger, e' possibile individuare
quali, tra le varie fonti di arrotondamento e imprecisione introdotte,
apportino un maggiore errore nella determinazione delle quantita'
fisiche finali.

Fonti di approssimazione presenti all'interno dell'apparato di trigger
sono:
\begin{itemize}
\item il campionamento del segnale effettuato in maniera non sincrona
      col tempo del picco in energia sul calorimetro elettromagnetico
\item la differenze nei fattori di amplificazione dei singoli canali
\item lo sfasamento relativo canale-canale
\item la compattazione delle celle in blocchetti $8\times2$
\item la digitalizzazione dell'informazione
\item la possibilita' di saturazione dei digitalizzatori ai vari livelli
\item il sistema di filtri con livello minimo di accettazione (soglia)
\item la somma digitale in righe e colonne di $128\times2$ celle
\item l'analisi nelle due proiezioni separate anziche' bidimensionale e
      di tutte le celle del calorimetro senza accorpamenti
\end{itemize}

Da numerosi test effettuati sull'apparato attraverso la sua simulazione
software emergono, in relazione alle diverse sezioni del Trigger neutro,
i seguenti errori relativi medi sulla ricostruzione dell'energia totale,
tutti distribuiti in maniera gaussiana, il cui effetto e' concorde
nell'apportare una minorazione dell'energia ricostruita dal trigger.

I risultati sono ottenuti da successive analisi, in ciascuna delle quali
vengono annullate tutte le possibili fonti di errore eccetto una.

{\samepage
\bce \btab{|l|l|}  \hline
fattore di amplificazione    & $-0.004\pm0.043$ \\
sfasamento canale-canale     & $-0.002\pm0.005$ \\
fase clock-picco             & $-0.102\pm0.075$ \\
ADC 10 bit                   & $-0.005\pm0.002$ \\
filtro temporale $0.5\s GeV$ & $-0.027\pm0.013$ \\
saturazione                  & $-0.000\pm0.002$ \\
floating point               & $-0.001\pm0.003$ \\
accorpamento celle           & $-0.000\pm0.000$ \\ \hline
\etab \ece
\addtocounter{tab}{1}
\centerline{Tab.\tab: Errori ed arrotondamenti}}
\vsp{0.5cm}

Il raggruppamento di piu' celle in gruppi non comporta, al fine di una
esatta ricostruzione dell'energia, alcun errore.

Escludendo i moduli Look-up table, la ricostruzione dell'energia totale
depositata sul calorimetro elettromagnetico avviene con un errore
relativo distribuito secondo una gaussiana con media $\mu=-0.132$,
$\sigma=0.147$, e troncata a 0.
L'energia ricostruita dal Peak-Sum risulta quindi inferiore all'energia
effettiva mediamente del $13\%$.

Solo in seguito ai moduli Look-up table, attraverso una ricostruzione
dell'energia totale ottenuta con interpolazione parabolica, l'errore
medio relativo sull'energia ricostruita rispetto all'energia
effettivamente depositata sul calorimetro risulta pari a
$-0.012\pm0.047$, con una conseguente diminuzione dell'errore di un
fattore 10 rispetto al precedente $-0.132\pm0.093$.

Le distribuzioni dell'energia e dell'errore relativo sull'energia sono
riportate, separatamente per i tre modi di decadimento \ksdpiz, \kldpiz,
\kltpiz, nelle prossime sezioni, insieme alle distribuzioni di numero di
gamma sul calorimetro, baricentro dell'energia, vertice di decadimento,
e errori relativi delle quantita' ricostruite rispetto alle
corrispondenti quantita' generate.

\section{Distribuzione temporale degli eventi}
\setcounter{fig}{0}
\setcounter{tab}{0}

\noindent
In funzione del tipo di analisi da effettuare, e' possibile, all'interno
del simulatore di Trigger neutro, selezionare diverse configurazioni per
la distribuzione temporale degli eventi.

E' possibile, ad esempio, tenere conto del branching ratio dei diversi
modi di decadimento neutri di \ks e \kl, cosa peraltro inutile ai fini
di una analisi della risposta dell'apparato ai tre diversi modi di
decadimento \ksdpiz, \kldpiz, \kltpiz, in quanto si verifica mediamente
un decadimento \kldpiz ogni 200 decadiment \kltpiz.

Constatato il fatto che la simulazione dell'undershoot non comporta
alcuna modifica apprezzabile ai risultati, essendo presente
nell'apparato di trigger un sottrattore di pie\-di\-stal\-lo (baseline
restoring) per ripristinare il livello di zero dopo gli ADC, gli eventi
vengono distribuiti in maniera tale da risultare temporalmente separati
in media di 10 slot temporali da $25\s ns$, tempo sufficiente per
garantire una non sovrapposizione dei segnali di shaper di due eventi
successivi, ed una corretta simulazione del flusso di dati attraverso i
vari stadi di pipeline.\\
La distribuzione temporale degli eventi puo' essere allargata o
ristretta a piacere, attraverso la modifica di alcuni parametri.

I test di cui sono riportati i risultati sono effettuati senza
sovrapposizione di eventi sulla stessa cella del calorimetro
elettromagnetico, fatto che peraltro si verifica con probabilita'
pressoche' nulla, dato il basso rate di fotoni sulla singola cella.

\section{Analisi e acquisizione dei dati}
\setcounter{fig}{0}
\setcounter{tab}{0}

\noindent
Una prima analisi dei dati generati dal programma Montecarlo consente di
ricostruire a livello del calorimetro elettromagnetico il numero di
gamma, l'energia rilasciata sul calorimetro, il baricentro dell'energia,
il vertice di decadimento del \k.
Tali valori vengono registrati come dati di livello 1.

I corrispondenti valori generati, quindi non ancora elaborati e
filtrati dal simulatore dell'apparato sperimentale, nonostante non
possano essere utilizzati dal trigger ne' da una successiva analisi
off-line perche' sconosciuti, vengono comunque registrati come dati di
livello 0.

In seguito i dati sono elaborati dal simulatore, e registrati in uscita
dal Trigger neutro come dati di livello 2, e come dati di livello 3 alla
fine dei moduli Look-up table.

I grafici ed i risultati riportati in seguito terranno quindi conto
della seguente notazione:

{\samepage
\bce \btab{|l|ll|}  \hline
livello 0: & NMC & New Montecarlo             \\
livello 1: & LKR & Liquid Krypton Calorimeter \\
livello 2: & NET & Neutral Trigger            \\
livello 3: & LUT & Look-up Table              \\ \hline
\etab \ece
\addtocounter{tab}{1}
\centerline{Tab.\tab: Notazione livelli}}
\vsp{0.5cm}

Per un raffronto dei dati ricostruiti a livello 1 e 3 fra loro e con i
dati generati di livello 0, e' essenziale numerare gli eventi e poter
disporre in uscita del numero dell'evento ricostruito.
Per questo scopo numerosi contatori sono disposti nel simulatore in
corrispondenza di ciascuno stadio di pipeline.

I dati a livello 1 e 3, ovvero ricostruiti da una lettura del
calorimetro elettromagnetico e dai moduli Look-up table rispettivamente,
sono riportati senza alcun taglio effettuato sui valori ricostruiti, al
fine di mettere in evidenza il reale funzionamento del trigger.

Solo in seguito, nel calcolo dell'efficienza dell'apparato sperimentale,
verranno effettuati i tagli necessari sulle varie quantita' fisiche
ricostruite.

I grafici riportati di seguito si riferiscono ad una analisi effettuata
su un campione di 10000 decadimenti per ognuno dei tre modi di
decadimento \ksdpiz, \kldpiz, \kltpiz, al fine di poter visualizzare i
risultati dei confronti tra le quantita' ai vari livelli.

\section{Energia}
\setcounter{fig}{0}
\setcounter{tab}{0}

\noindent
Le distribuzioni in energia, per i tre modi di decadimento analizzati,
sono graficate con l'energia totale, espressa in $GeV$, sulle ascisse, e
il numero di eevnti con energia compresa nell'imtervallo di
$\pm0.5\s GeV$ attorno al valore del singolo step dell'istogramma sulle
ordinate, e si riferiscono nell'ordine a:
\begin{itemize}
\item Energia generata dal programma Montecarlo (NMC): ENERG 0
\item Energia ricostruita da Look-up Table (LUT): ENERG 3
\end{itemize}

I grafici di raffronto si riferiscono al plot dei valori generati o
ricostruiti ai diversi livelli per eventi corrispondenti;
il primo indice si riferisce al valore riportato sulle ascisse, il
secondo indice al valore riportato sulle ordinate del grafico.

\bce
\addtocounter{fig}{1}
Fig.\fig: \bksdpiz Energia generata NMC e ricostruita LUT
\ece
\vsp{0.5cm}

\bce
\addtocounter{fig}{1}
Fig.\fig: \bksdpiz Energia ricostruita LUT vs. generata NMC
\ece
\vsp{0.5cm}

Gli eventi che, graficati, si trovano in prossimita' della retta a
$45^\circ$ sono quegli eventi per cui tutta l'energia del \ks
viene depositata sul calorimetro elettromagnetico.
I punti invece che si trovano al di sotto della retta a $45^\circ$ sono
riferiti a quegli eventi per cui uno o piu' fotoni non giungono sul
calorimetro: la deposizione di energia risulta quindi inferiore a
quella posseduta dalla particella originaria.

\bce
\addtocounter{fig}{1}
Fig.\fig: \bkldpiz Energia generata NMC e ricostruita LUT
\ece
\vsp{0.5cm}

\bce
\addtocounter{fig}{1}
Fig.\fig: \bkldpiz Energia ricostruita LUT vs. generata NMC
\ece
\vsp{0.5cm}

La differente distribuzione in energia per i decadimenti \ksdpiz da una
parte, e \kldpiz, kltpiz dall'altra, e' dovuta essenzialmente alla
configurazione dell'apparato sperimentale e al sistema di produzione dei
due fasci \ks e \kl rispettivamente.

\bce
\addtocounter{fig}{1}
Fig.\fig: \bkltpiz Energia generata NMC e ricostruita LUT
\ece
\vsp{0.5cm}

\bce
\addtocounter{fig}{1}
\nopagebreak
Fig.\fig: \bkltpiz Energia ricostruita LUT vs. generata NMC
\ece
\vsp{0.5cm}

Dall'analisi della distribuzione dell'energia ricostruita a livello 3
(LUT), e dal successivo raffronto con l'energia generata a livello 0
(NMC) per i tre modi di decadimento, e' possibile notare come per un
vasto numero di eventi l'energia ricostruita sia inferiore a quella
generata; questo effetto e' dovuto agli eventi per cui uno o piu' gamma
non giungono sul calorimetro elettromagnetico, senza quindi che possa
essere rivelata la loro energia.

Un successivo raffronto tra l'energia ricostruita a livello 1 (LKR)
sulle ascisse e a livello 3 (LUT) sulle ordinate mette invece in risalto
la precisione dell'apparato di Trigger e Look-up table nella
ricostruzione dell'energa totale depositata (grafici di sinistra).

Attraverso il calcolo
\beq \frac{Energia(3)-Energia(1)}{Energia(1)} \eeq
e' possibile stimare l'errore relativo nella ricostruzione dell'energia
totale da parte del Trigger neutro (grafici di destra).

\bce
\addtocounter{fig}{1}
Fig.\fig: \bksdpiz Energia 1-3 e errore relativo
\ksdpiz \hsp{1cm} $errore=-0.017\pm0.038$
\ece
\vsp{0.5cm}

\bce
\addtocounter{fig}{1}
\nopagebreak
Fig.\fig: \bkldpiz Energia 1-3 e errore relativo
\nopagebreak
\kldpiz \hsp{1cm} $errore=-0.013\pm0.046$
\ece
\vsp{0.5cm}

\bce
\addtocounter{fig}{1}
Fig.\fig: \bkltpiz Energia 1-3 e errore relativo
\kltpiz \hsp{1cm} $errore=-0.012\pm0.049$
\ece
\vsp{0.5cm}

\section{Numero di gamma}
\setcounter{fig}{0}
\setcounter{tab}{0}

\noindent
Mentre i gamma generati sono ovviamente 4 per i decadimenti \ksdpiz,
\kldpiz, e 6 per i decadimenti \kltpiz, i gamma che in realta'
colpiscono il calorimetro elettromagnetico possono essere in numero
inferiore, a causa di quei gamma che hanno traiettorie esterne al
rivelatore e quindi anticoincisi, o che, poco energetici, non vengono
rivelati.

Per ognuno dei tre modi di decadimento sono riportate, nell'ordine, le
seguenti di\-stri\-bu\-zioni:

\begin{itemize}
\item Numero di gamma che giungono sul calorimetro elettromagnetico:
      GAMMA 0
\item Numero di gamma ricostruiti sul calorimetro elettromagnetico da
      una analisi cella per cella, con una soglia per l'individuazione
      del picco fissata a $0.5\s GeV$ ed un algoritmo di ricerca che
      prevede il confronto di ogni singola cella con le otto
      circostanti: GAMMA 1
\item Numero di gamma ricostruiti sul calorimetro dall'analisi nella
      proiezione $x$, senza alcun taglio effettuato, delle colonne gia'
      accorpate a due a due: GAMMA 1 X, in modo da rendere questa
      distribuzione confrontabile con quella ottenuta a livello 3.
      La soglia per l'individuazione di un picco in una colonna e'
      fissata a $0.5\s GeV$, valore corrispondente alla soglia del
      filtro temporale, e l'algoritmo di ricerca spaziale prevede il
      confronto di ogni colonna con le due adiacenti (per le due colonne
      esterne il confronto avviene con uno zero)
\item Differenza tra il numero di picchi rivelati dal trigger (livello
      3) e il numero di gamma ricostruiti sul calorimetro
      elettromagnetico dall'analisi delle celle gia' accorpate in
      colonne (livello 1) nella proiezione $x$: GAMMA 3-1 X
\end{itemize}

\bce
\addtocounter{fig}{1}
Fig.\fig: \bksdpiz Numero gamma generati e ricostruiti cella per cella
\ece
\vsp{0.5cm}

\bce
\addtocounter{fig}{1}
Fig.\fig: \bksdpiz Numero gamma in proiezione $x$ e errore di
                   ricostruzione
\ece
\vsp{0.5cm}

Per i decadimenti \ksdpiz, il numero di eventi in cui almeno un gamma
giunge sul calorimetro e' pari al $62.4\%$ del totale dei decadimenti
\ksdpiz prodotti.
Tale riduzione e' dovuta essenzialmente ai collimatori e alla
anticoincidenza anti-\ks, impiegata per sopprimere i decadimenti
\ksdpiz con fotoni esterni al cono di accettanza del rivelatore, e
quindi anticoincisi.

\vsp{0.5cm}

\bce
\addtocounter{fig}{1}
Fig.\fig: \bkldpiz Numero gamma generati e ricostruiti cella per cella
\ece
\vsp{0.5cm}

\bce
\addtocounter{fig}{1}
Fig.\fig: \bkldpiz Numero gamma in proiezione $x$ e errore di
                   ricostruzione
\ece
\vsp{0.5cm}

\vsp{1.0cm}

\bce
\addtocounter{fig}{1}
Fig.\fig: \bkltpiz Numero gamma generati e ricostruiti cella per cella
\ece
\vsp{0.5cm}

\bce
\addtocounter{fig}{1}
Fig.\fig: \bkltpiz Numero gamma in proiezione $x$ e errore di
                   ricostruzione
\ece
\vsp{0.5cm}

Per i decadimenti \kldpiz e \kltpiz il numero di eventi in cui almeno un
gamma giunge sul calorimetro elettromagnetico e' pari a circa il $99\%$
del totale dei decadimenti prodotti all'interno della zona fiduciale, in
quanto non esiste un anticoincidenza anti-\kl subito oltre la targhetta
\kl.

E' evidente come nelle analisi per colonne di $128\times2$ celle,
effettuate a livello del calorimetro elettromagnetico e a livello del
Trigger neutro, si abbia una diminuzione del numero di eventi rivelati a
4(6) gamma, rispetto al numero di eventi generati a 4(6) gamma.
Tale effetto e' dovuto a eventi in cui piu' gamma colpiscono il
calorimetro elettromagnetico in un intervallo spaziale orizzontale di
$4\s cm$, corrispondente alla larghezza di 2 celle, ovvero di una
colonna, venendo quindi rivelati come un singolo picco all'interno della
medesima colonna nella proiezione $x$.

Un numero di gamma ricostruito superiore al numero di gamma incidenti
sul calorimetro e' invece dovuto o ad effetti di sovrapposizione di due
o piu' sciami elettromagnetici di differenti gamma, che generano quindi
dei punti di sella nella distribuzione spaziale dell'energia, o ad
effetti di bordo, consistenti nel confronto delle due colonne esterne
con la rispettiva colonna interna adiacente da una parte, con uno zero
dall'altra, algoritmo questo utilizzato dal Peak-Finder per
l'individuazione dei picchi.

Per la proiezione $y$ i risultati e i grafici sono pressoche' identici.

\section{Baricentro dell'energia}
\setcounter{fig}{0}
\setcounter{tab}{0}

\noindent
Il baricentro dell'energia depositata sul calorimetro elettromagnetico
e' calcolato, separatamente per le proiezioni $x$ e $y$, attraverso la
formula
\beq M_{1X}=\frac{\sum E_i X_i}{\sum E_i} \eeq
\beq M_{1Y}=\frac{\sum E_i Y_i}{\sum E_i} \eeq
come momento di ordine 1 dell'energia rispetto alla posizione
proiettata.

I plot del baricentro dell'energia, espresso in $cm$, per i tre modi di
decadimento, si riferiscono, nell'ordine, a:

\begin{itemize}
\item Plot cartesiano del baricentro, calcolato dal punto di impatto dei
      gamma generati sul calorimetro elettromagnetico: BARIC 0
\item Distribuzione della distanza del baricentro dell'energia,
      calcolato a partire dalla conoscenza del punto d'impatto dei
      singoli gamma sul calorimetro, dal centro del calorimetro
      elettromagnetico: BARIC 0
\item Plot cartesiano del baricentro dell'energia depositata
      sul calorimetro ricostruito dal Trigger neutro: BARIC 3
\item Distribuzione della distanza del baricentro dell'energia,
      ricostruito dal Trigger neutro, dal centro del calorimetro
      elettromagnetico: BARIC 3
\end{itemize}

\bce
\addtocounter{fig}{1}
Fig.\fig: \bksdpiz Baricentro generato
\ece
\vsp{0.5cm}

\bce
\addtocounter{fig}{1}
Fig.\fig: \bksdpiz Baricentro ricostruito
\ece
\vsp{0.5cm}


Nella ricostruzione del baricentro dell'energia depositata sul
calorimetro elet\-tro\-ma\-gne\-ti\-co, occorre tenere presente che la
posizone cartesiana dei gamma non e' nota, bensi' se ne conoscono le due
posizioni proiettate sulle viste $x$ e $y$, con una precisione
dell'ordine di $4\s cm$, corrispondenti alla larghezza di ciascun
canale.

\bce
\addtocounter{fig}{1}
Fig.\fig: \bkldpiz Baricentro generato
\ece
\vsp{0.5cm}

\bce
\addtocounter{fig}{1}
Fig.\fig: \bkldpiz Baricentro ricostruito
\ece
\vsp{0.5cm}

\vsp{1.0cm}

\bce
\addtocounter{fig}{1}
Fig.\fig: \bkltpiz Baricentro generato
\ece
\vsp{0.5cm}

\bce
\addtocounter{fig}{1}
Fig.\fig: \bkltpiz Baricentro ricostruito
\ece
\vsp{0.5cm}

Attraverso il calcolo
\beq \frac{Baricentro(3)-Baricentro(1)}{Baricentro(1)} \eeq
e' possibile stimare l'errore relativo sul baricentro dell'energia
ricostruito dal Trigger neutro in funzione dello stesso ricostruito
dall'analisi delle singole celle del calorimetro elettromagnetico
accorpate in righe e colonne $128\times2$.\\
L'errore nella ricostruzione risulta essere, per i tre modi di
decadimento: BARIC 3/1.

\bce
\addtocounter{fig}{1}
Fig.\fig: \bksdpiz Errore sulla ricostruzione del baricentro
\ksdpiz \hsp{1cm} $errore=0.004\pm0.083$
\ece
\vsp{0.5cm}

\bce
\addtocounter{fig}{1}
Fig.\fig: \bkldpiz Errore sulla ricostruzione del baricentro
\kldpiz \hsp{1cm} $errore=0.018\pm0.101$
\ece
\vsp{0.5cm}

\bce
\addtocounter{fig}{1}
Fig.\fig: \bkltpiz Errore sulla ricostruzione del baricentro
\kltpi\ \hsp{1cm} $errore=0.021\pm0.196$
\ece
\vsp{0.5cm}

\section{Vertice di decadimento della particella}
\setcounter{fig}{0}
\setcounter{tab}{0}

\noindent
I plot per i tre modi di decadimento del vertice di decadimento della
particella \ks o \kl, espresso in $cm$, si riferiscono nell'ordine a:

\begin{itemize}
\item Distribuzione del vertice di decadimento del \k generato da NMC
      nella zona fiduciale: VERTX 0
\item Raffronto tra vertice generato da NMC a livello 0 (sulle ascisse)
      e ricostruito sul calorimetro elettromagnetico a livello 1 (sulle
      ordinate): VERTX 0-1
\item Distribuzione del vertice di decadimento del \k ricostruito dal
      Trigger neutro senza tagli effettuati: VERTX 3
\item Raffronto tra vertice generato da NMC a livello 0 (sulle ascisse)
      e vertice ricostruito dal trigger a livello 3 (sulle ordinate):
      VERTX 0-3
\end{itemize}

\bce
\addtocounter{fig}{1}
Fig.\fig: \bksdpiz Vertice generato e raffronto NMC-LKR
\ece
\vsp{0.5cm}

\bce
\addtocounter{fig}{1}
Fig.\fig: \bksdpiz Vertice ricostruito e raffronto NMC-NET
\ece
\vsp{0.5cm}

Il vertice generato e ricostruito per i decadimenti \ks presenta un
taglio inferiore abbastanza netto in corrispondenza dell'anticoincidenza
e dell'ultimo collimatore \ks, a $6\s m$ dalla targhetta \ks;
attraverso il foro del collimatore riescono comunque a passare i gamma
prodotti da alcuni decadimenti di \ks precedenti al collimatore stesso,
il che giustifica una distribuzione non nulla prima dei $6\s m$.\\

\vsp{1.0cm}

\bce
\addtocounter{fig}{1}
Fig.\fig: \bkldpiz Vertice generato e raffronto NMC-LKR
\ece
\vsp{0.5cm}

\bce
\addtocounter{fig}{1}
Fig.\fig: \bkldpiz Vertice ricostruito e raffronto NMC-NET
\ece
\vsp{0.5cm}

\vsp{1.0cm}

\bce
\addtocounter{fig}{1}
Fig.\fig: \bkltpiz Vertice generato e raffronto NMC-LKR
\ece
\vsp{0.5cm}

\bce
\addtocounter{fig}{1}
Fig.\fig: \bkltpiz Vertice ricostruito e raffronto NMC-NET
\ece
\vsp{0.5cm}

Il vertice generato e ricostruito per i decadimenti \kldpiz e \kltpiz
presenta una distribuzione non nulla per distanze dall'origine
del sistema di riferimento inferiori a $4.8\s m$, dove e' posizionato
l'ultimo collimatore \kl;
attraverso il foro del collimatore riescono a passare i gamma prodotti
da decadimenti di alcuni \kl precedenti al collimatore stesso, il che
giustifica una distribuzione non nulla prima di $4.8\s m$.\\

Dall'analisi delle distribuzioni del vertice di decadimento ricostruito
a livello 3 (NET) e dal successivo raffronto col vertice generato a
livello 0 (NMV), per i tre modi di decadimento, e' possibile notare come
per un vasto numero di eventi il vertice di decadimento ricostruito
risulti superiore a quello generato.
Questo e' dovuto al fatto che per gli eventi in cui uno o piu' gamma non
giungono sul calorimetro, il vertice di decadimento, in accordo con le
previsioni, viene ricostruito ad una distanza maggiore dalla targhetta
\ks, origine del sistema di riferimento adottato.\\
La formula utilizzata infatti per il calcolo del vertice di decadimento
della particella e':
\beq Z_V=D-\frac{E}{m_K}\sqrt{(M_{2X}+M_{2Y})-(M_{1X}^2+M_{1Y}^2)} \eeq
dove $Z_V$ e' la distanza del vertice di decadimento dalla targhetta
\ks, $D$ la distanza di tale targhetta dal calorimetro elettromagnetico
($122.35\s m$), $E$ l'energia totale ricostruita, $m_k$ la massa del \k
($0.497\s GeV$), $M_{1X,Y}$ e $M_{2X,Y}$ rispettivamente i momenti
dell'energia di ordine 1 e 2 rispetto a $X,Y$.\\
La formula per il calcolo del vertice di decadimento $Z_V$ e' data dalla
conservazione del quadriimpulso totale del sistema, nell'approssimazione
di piccoli angoli, e dalla conoscenza della massa invariante del $K$.

Un successivo raffronto tra il vertice di decadimento ricostruito a
livello 3 (NET) e a livello 1 (LKR) mette in risalto la precisione
dell'apparato di Trigger e Look-up table nella ricostruzione del vertice
(grafici di sinistra).

Attraverso il calcolo
\beq \frac{Vertice(3)-Vertice(1)}{Vertice(1)} \eeq
e' possibile stimare l'errore relativo nella ricostruzione del vertice
di decadimento da parte dell'elettronica (grafici di destra).

\bce
\addtocounter{fig}{1}
Fig.\fig: \bksdpiz Vertice 1-3 e errore relativo
\ksdpiz \hsp{1cm} $errore=0.027\pm0.041$
\ece
\vsp{0.5cm}

\bce
\addtocounter{fig}{1}
Fig.\fig: \bkldpiz Vertice 1-3 e errore relativo
\ksdpiz \hsp{1cm} $errore=0.042\pm0.079$
\ece
\vsp{0.5cm}

\bce
\addtocounter{fig}{1}
Fig.\fig: \bkltpiz Vertice 1-3 e errore relativo
\ksdpiz \hsp{1cm} $errore=0.051\pm0.107$
\ece
\vsp{0.5cm}

\noindent
In tutti i grafici e' possibile notare una  forte asimmetria tendente ad
una ricostruzione del vertice di decadimento superiore a quello
effettivo, come peraltro previsto.

\section{Efficienza del Trigger neutro}
\setcounter{fig}{0}
\setcounter{tab}{0}

\noindent
In funzione di diversi parametri e di diverse configurazioni
dell'apparato sperimentale si ottengono differenti efficienze nella
rivelazione dei tre modi di decadimento \ksdpiz, \kldpiz, \kltpiz.

Possibili variazioni alle soglie dei filtri temporali, un differente
sistema di raggruppamento delle celle in blocchetti prima, in righe e
colonne poi, e diversi tagli effettuati sui valori ricostruiti
comportano una diversa efficienza del sistema.

Unitamente all'efficienza di rivelazione del decadimento, e' necessario
considerare anche il numero di eventi che costituiscono il fondo
\kltpiz rivelati dal trigger come eventi di decadimenti \kldpiz, al fine
di ottimizzare l'acquisizione dati in funzione di questi due parametri.

Un intervallo di accettazione troppo ampio per il vertice di decadimento
ricostruito, mentre da una parte consentirebbe una elevata efficienza
nella rivelazione dei decadimenti  \ksdpiz e \kldpiz puliti, dall'altra
consentirebbe a parecchi decadimenti \kltpiz di essere rivelati come
decadimenti \kldpiz.\\
Dal momento che il branching ratio dei decadimenti \kltpiz e' 0.217, e
il branching ratio dei decadimenti \kldpiz e' 0.001, e quindi in natura
avviene un decadimento \kldpiz ogni 210 decadimenti \kltpiz, una
inefficienza nel riconoscimento dei decadimenti \kltpiz comporterebbe
la registrazione su supporto magnetico di un elevato numero di eventi
indesiderati.

La conoscenza dell'efficienza con una precisione di circa
$1\cdot10^{-4}$ e' necessaria per una misura del doppio rapporto
\beq R=\frac{N(K_L\rightarrow\pi^0\pi^0)}{N(K_S\rightarrow\pi^0\pi^0)}
/ \frac{N(K_L\rightarrow\pi^+\pi^-)}{N(K_S\rightarrow\pi^+\pi^-)}
\approx 1-6Re(\frac{\epsilon\prime}{\epsilon}) \eeq
con una precisione di $2\cdot10^{-4}$.\\
Detto infatti $n$ il numero reale di decadimenti di un certo tipo, e
$n_0$ il numero misurato di tali decadimenti, dalla conoscenza
dell'efficienza $e$ relativa a quel modo di decadimento, e' possibile
ricostruire il numero $n_1$ di eventi stimati come:
\beq n_1=\frac{n_0}{e} \eeq

Per una accurata misura di $R$ occorre quindi stimare con la precisione
richiesta $n_1$ per i quattro modi di decadimento \ksdpic, \ksdpiz,
\kldpic, \kldpiz, ovvero l'efficienza $e$, dal momento che $n_0$ risulta
dall'esperimento.

Essendo la distribuzione degli eventi nell'unita' di tempo di tipo
Poissoniano, caratte\-riz\-za\-ta da media $\mu=n$, con
$n$ numero di eventi, e $\sigma=\sqrt{\mu}$, ovvero da errore relativo
\beq \frac{\sigma}{\mu}=\frac{1}{\sqrt{\mu}} \eeq
il numero di decadimenti necessari per avere una precisione sul doppio
rapporto dell'ordine di $10^{-4}$ e' circa $10^8$.

Nelle seguenti tabelle sono riportati i rapporti tra il numero di eventi
in uscita (out), e il numero totale di eventi generati da NMC (tot) e in
ingresso al Trigger (in), in funzione di differenti configurazioni
possibili dell'apparato sperimentale di Trigger neutro.
I valori 1 e 0 si riferiscono rispettivamente alla accettazione o meno
dell'evento.

Gli eventi in ingresso (in) sono accettati se si presentano come
decadimenti \ksdpiz o \kldpiz con 4 gamma sul calorimetro
elettromagnetico (il dato e' fornito dal programma di generazione degli
eventi, e, benche' sconosciuto nella realta', e' utile a questo livello
per una esatta conoscenza dell'efficienza dell'apparato).\\
Gli eventi in uscita (out) sono accettati se giudicati buoni dal Trigger
neutro, ovvero se, una volta filtrati ed elaborati dall'apparato, i
valori in uscita rientrano entro gli intervalli di accettazione
richiesti.

Il rapporto $out1/in1$ esprime direttamente l'efficienza nel conteggio
degli eventi buoni, mentre il rapporto $out0/in0$ esprime l'efficienza
nell'eliminazione del fondo e dei decadimenti non buoni.\\
I rapporti incrociati $out1/in0$ e $out0/in1$ tengono conto della
inefficienza dell'apparato di trigger nell'eliminazione del fondo e
nella rivelazione di eventi buoni rispettivamente.

Il rapporto $K_L2/K_L3$ indica ogni quanti eventi registrati costituenti
il fondo \kltpiz viene registrato un decadimento utile \kldpiz, tenuto
conto del branching ratio dei due modi dei decadimento.
Tale indicazione risulta estremamente importante per la mole di
informazioni e il rate di eventi da registrare su disco.

I valori di seguito riportati si riferiscono ad una analisi in assenza
di eventi accidentali in funzione di differenti configurazioni e tagli
effettuati per l'accettazione delle quantita' fisiche ricostruite.

Configurazione: $blocchetti$ indica il numero di celle componenti
ciascun blocchetto e la loro configurazione, $soglia$ e' la soglia del
filtro temporale espressa in $GeV$.

Tagli: gli intervalli di accettazione per le quattro quantita' fisiche
ricostruite esprimono il valore minimo e massimo rispettivamente per
l'accettazione dell'evento in esame;
$gamma$ e' il numero di gamma nella singola proiezione,
$energia$ e' l'energia totale ricostruita del \k espressa in $GeV$,
$baricentro$ e' la distanza, espressa in $cm$, del baricentro
dell'energia ricostruito dal centro del calorimetro elettromagnetico,
$vertice$ e' il vertice di decadimento della particella espresso in
$cm$.\\

\bce \btab{|cc|} \hline
blocchetti  & $8\times2$   \\
soglia      & 0.5          \\ \hline
energia     & (60,180)     \\
baricentro  & (0,10)       \\ \hline
\etab \ece
\addtocounter{tab}{1}
\centerline{Tab.\tab: Configurazione e tagli}
\vsp{0.5cm}

I tagli su energia e baricentro dell'energia sono dedotti dall'analisi
delle distribuzioni delle suddette quantita' generate e ricostruite.
Per quanto riguarda invece numero di gamma e vertice di decadimento, in
funzione dei differenti tagli applicati, si ottengono differenti valori
di efficienza e quindi di rate di eventi da registrare su supporto
magnetico.\\

\bce \btab{|lcc|c|c|c|} \hline
gamma         &     &     &   3,4    &   2,5    &   3,4    \\
vertice       &     &     & 300,1500 & 200,2000 & 200,2000 \\ \hline
              & in  & out &          &          &          \\ \hline
\bksdpiz      &  0  &  0  &   0.997  &   0.962  &   0.890  \\
              &  0  &  1  &   0.003  &   0.038  &   0.110  \\
in0/tot=0.666 &  1  &  0  &   0.641  &   0.244  &   0.521  \\
in1/tot=0.334 &  1  &  1  &   0.359  &   0.756  &   0.479  \\
              & tot &  1  &   0.112  &   0.278  &   0.167  \\ \hline
\bkldpiz      &  0  &  0  &   0.996  &   0.972  &   0.992  \\
              &  0  &  1  &   0.004  &   0.028  &   0.008  \\
in0/tot=0.603 &  1  &  0  &   0.854  &   0.615  &   0.774  \\
in1/tot=0.397 &  1  &  1  &   0.146  &   0.385  &   0.226  \\
              & tot &  1  &   0.060  &   0.170  &   0.095  \\ \hline
\bkltpiz      &  0  &  0  &   0.991  &   0.939  &   0.985  \\
              &  0  &  1  &   0.009  &   0.061  &   0.015  \\
in0/tot=1.000 &  1  &  0  &   0.000  &   0.000  &   0.000  \\
in1/tot=0.000 &  1  &  1  &   0.000  &   0.000  &   0.000  \\
              & tot &  1  &   0.009  &   0.061  &   0.015  \\ \hline
$K_L2/K_L3$   &     &     &   1/31   &   1/75   &    1/33  \\ \hline
\etab \ece
\addtocounter{tab}{1}
\centerline{Tab.\tab: Efficienza}
\vsp{0.5cm}

I valori riportati si riferiscono ad una analisi effettuata con tagli in
ricostruzione relativamente stretti sull'accettazione del vertice di
decadimento della particella \ks o \kl.\\
Cosi' facendo e' possibile ridurre il fondo dei decadimenti \kltpiz, a
scapito comunque dell'efficienza nel conteggio dei decadimenti \ksdpiz e
\kldpiz.

\bce \btab{|lcc|c|c|c|} \hline
gamma         &     &     &   3,4    &   3,4    &     2,5    \\
vertice       &     &     & 200,1000 & 200,5000 & -1000,5000 \\ \hline
              & in  & out &          &          &            \\ \hline
\bksdpiz      &  0  &  0  &   0.999  &   0.923  &   0.783  \\
              &  0  &  1  &   0.001  &   0.077  &   0.217  \\
in0/tot=0.666 &  1  &  0  &   0.820  &   0.337  &   0.098  \\
in1/tot=0.334 &  1  &  1  &   0.180  &   0.663  &   0.902  \\
              & tot &  1  &   0.061  &   0.273  &   0.446  \\ \hline
\bkldpiz      &  0  &  0  &   0.999  &   0.908  &   0.741  \\
              &  0  &  1  &   0.001  &   0.092  &   0.259  \\
in0/tot=0.603 &  1  &  0  &   0.921  &   0.344  &   0.104  \\
in1/tot=0.397 &  1  &  1  &   0.079  &   0.656  &   0.896  \\
              & tot &  1  &   0.032  &   0.316  &   0.512   \\ \hline
\bkltpiz      &  0  &  0  &   0.000  &   0.705  &   0.659  \\
              &  0  &  1  &   0.005  &   0.295  &   0.341  \\
in0/tot=1.000 &  1  &  0  &   0.000  &   0.000  &   0.000  \\
in1/tot=0.000 &  1  &  1  &   0.000  &   0.000  &   0.000  \\
              & tot &  1  &   0.005  &   0.295  &   0.341  \\ \hline
$K_L2/K_L3$   &     &     &   1/33   &   1/196  &    1/139 \\ \hline
\etab \ece
\addtocounter{tab}{1}
\nopagebreak
\centerline{Tab.\tab: Efficienza}
\vsp{0.5cm}

I valori riportati si riferiscono ad eventi per cui in almeno una cella
del calorimetro elettromagnetico venga rilasciata un'energia non
nulla.\\
Per i decadimenti \ksdpiz occorre tenere conto che solo il $61\%$ dei
decadimenti generati arrivano sul calorimetro elettromagnetico senza
essere bloccati dall'anticoincidenza \ks.\\
Per i decadimenti \kldpiz e \kltpiz occorre considerare che gli eventi
analizzati si riferiscono ad una piccola parte degli eventi
effettivamente presenti nella realta', in quanto gia' a livello di
generazione da parte del Montecarlo vengono creati soltanto eventi per
cui il vertice di decadimento rientra in un intervallo di
($-100,3500\s cm$) dalla targhetta \ks.

Per il calcolo del doppio rapporto $R$ da una successiva analisi
off-line verranno presi in considerazione soltanto decadimenti \ksdpiz
e \kldpiz per i quali energia e vertice di decadimento rientrano entro
determinati intervalli di accettazione.
Cosi' facendo si puo' ovviare all'inconveniente di non conoscere
esattamente la totalita' degli eventi di decadimento utili alla misura
di $R$.

Tali risultati suggeriscono, al fine di una buona efficienza
nell'acquisizione dei decadimenti \ksdpiz, \kldpiz, senza una eccessiva
accettanza del fondo \kltpiz, rispetto ai tagli proposti \cite{B317}, un
restringimento dell'intervallo di accettazione per il vertice di
decadimento del \k, e un allargamento dell'intervallo di accettazione
per il numero di gamma visti nella singola proiezione sul calorimetro
elettromagnetico.\\

\bce \btab{|c|c|c|} \hline
        & proposta   & modifica   \\ \hline
gamma   & (3,4)      & (2,5)      \\
vertice & (200,5000) & (200,2000) \\ \hline
\etab \ece
\addtocounter{tab}{1}
\centerline{Tab.\tab: Tagli: proposta e modifica}
\vsp{0.5cm}

\subsection{Variazione di soglia}

\noindent
Una soglia di $1.0\s GeV$ per il filtro temporale implica una
diminuzione dell'efficienza totale nell'accettazione degli eventi buoni
dello $0.53\%$ per i modi di decadimento \ksdpiz e \kldpiz, mentre
l'inefficienza nell'accettazione dei decadimenti \kltpiz si riduce del
$2.17\%$.

\subsection{Variazione di raggruppamento delle celle}

\noindent
Variando il raggruppamento in blocchetti delle singole celle, e passando
ad una configurazione in cui ciascun blocchetto e' formato da
$16\times2$ celle anziche' da $8\times2$ celle, nel caso in cui il
valore della soglia del filtro temporale venga posto a $1.0 \s GeV$,
l'efficienza totale nell'accettazione degli eventi buoni cala del
$1.37\%$ per i modi di decadimento \ksdpiz e \kldpiz, mentre
l'inefficienza nell'accettazione dei decadimenti \kltpiz si riduce del
$1.52\%$.

\section{Eventi accidentali}
\setcounter{fig}{0}
\setcounter{tab}{0}

\noindent
Per una accurata simulazione della realta' fisica occorre prendere in
considerazione, oltre alle fonti di rumore interne all'apparato, anche
eventuali fenomeni che, dall'esterno, possono disturbare o sporcare
l'evento in esame.

Tali fenomeni, comunemente indicati come eventi accidentali,
costituiscono, insieme alla rimozione del fondo \kltpiz, il grosso
problema a livello di una analisi dei dati a qualsiasi livello.\\
Un decadimento ad esempio con soli tre gamma rivelati sul calorimetro
elettromagnetico, in coincidenza del quale sia presente un gamma
accidentale, puo' essere rivelato dall'apparato come un evento a quattro
gamma, e, quantomeno a livello di numero di picchi ricostruiti,
candidato ad essere un evento \ksdpiz o \kldpiz, quindi da registrare.

Attraverso una analisi dei decadimenti \ksdpiz, \kldpiz, \kltpiz in
presenza di eventi accidentali, siano essi dovuti a gamma esterni
particolarmente energetici non schermati dall'apparato, siano essi
frutto di altri modi di decadimento del $K$, e' possibile verificare
la variazione di efficienza del Trigger neutro rispetto ad una
condizione ideale di funzionamento precedentemente trattata.

Una analisi del comportamento del Trigger neutro in presenza di eventi
esterni e' effettuata con un rate di gamma accidentali pari al rate di
decadimento di \ks e \kl che decadono in zona fiduciale, ed in modo tale
da sincronizzare il gamma accidentale con la finestra del picco in
energia rilasciata sul calorimetro formato dagli shapers.

Per la distribuzione di tali eventi accidentali sono stati creati
appositi moduli Fortran che, avvalendosi del metodo Montecarlo,
generano un gamma con energia, impulso, punto d'impatto e tempo di
impatto sul calorimetro elettromagnetico determinati.\\
Mentre per la distribuzione del punto di impatto sul calorimetro e'
prevista una di\-stri\-bu\-zio\-ne casuale uniforme su tutta la sua
superficie, per quanto riguarda l'energia del gamma accidentale, questa,
sperimentalmente misurata nel corso dei test effettuati su NA48, risulta
distribuita con valor medio $\mu=20\s GeV$, sigma $\sigma=15\s GeV$,
asimmetria con skewness pari a 2.

L'interazione del gamma col Krypton liquido, e la conseguente produzione
dello sciame elettromagnetico, vengono simulate, ai fini del calcolo
dell'energia rilasciata su ciascuna cella del calorimetro, attraverso
l'utilizzo di una distribuzione dell'energia in funzione della distanza
dal centro dello sciame.
Tale distribuzione, nota sperimentalmente e funzione dell'energia del
gamma incidente, risulta dalla sovrapposizione di tre esponenziali
de\-cre\-scen\-ti in funzione della distanza e corrispondenti
rispettivamente a testa, corpo, coda del singolo sciame
elettromagnetico.

Il tempo di arrivo del gamma accidentale, in questo tipo di simulazione,
viene scelto in modo tale da cadere, con distribuzione casuale uniforme,
all'interno della finestra temporale del picco in energia
rilasciata sul calorimetro elettromagnetico, e corrispondente a
$110\s ns$.

Con la convenzione precedentemente adottata, ovvero indicando con $in0$
e $in1$ rispettivamente gli eventi in ingresso non buoni e buoni in
assenza di accidentali, e con $out0$ e $out1$ gli eventi in uscita
giudicati non buoni e buoni in presenza di accidentali, si ottengono
per i tre modi di decadimento i seguenti risultati:\\

\bce \btab{|cc|} \hline
blocchetti  & $8\times2$   \\
soglia      & 0.5          \\ \hline
energia     & (60,180)     \\
baricentro  & (0,10)       \\ \hline
\etab \ece
\addtocounter{tab}{1}
\nopagebreak
\centerline{Tab.\tab: Configurazione  e tagli}

\bce \btab{|lcc|c|c|c|} \hline
gamma         &     &     &   3,4    &   2,5    &     2,5    \\
vertice       &     &     & 200,5000 & 200,2000 & -1000,5000 \\ \hline
              & in  & out &          &          &            \\ \hline
\bksdpiz      &  0  &  0  &   0.916  &   0.984  &   0.886  \\
              &  0  &  1  &   0.084  &   0.016  &   0.114  \\
in0/tot=0.666 &  1  &  0  &   0.468  &   0.743  &   0.373  \\
in1/tot=0.334 &  1  &  1  &   0.532  &   0.257  &   0.627  \\
              & tot &  1  &   0.234  &   0.096  &   0.285  \\ \hline
\bkldpiz      &  0  &  0  &   0.893  &   0.977  &   0.807  \\
              &  0  &  1  &   0.107  &   0.023  &   0.193  \\
in0/tot=0.603 &  1  &  0  &   0.378  &   0.830  &   0.217  \\
in1/tot=0.397 &  1  &  1  &   0.622  &   0.170  &   0.783  \\
              & tot &  1  &   0.312  &   0.081  &   0.427  \\ \hline
\bkltpiz      &  0  &  0  &   0.698  &   0.977  &   0.559  \\
              &  0  &  1  &   0.302  &   0.023  &   0.441  \\
in0/tot=1.000 &  1  &  0  &   0.000  &   0.000  &   0.000  \\
in1/tot=0.000 &  1  &  1  &   0.000  &   0.000  &   0.000  \\
              & tot &  1  &   0.302  &   0.023  &   0.441  \\ \hline
\etab \ece
\addtocounter{tab}{1}
\centerline{Tab.\tab: Efficienza in presenza di accidentali intensi}
\vsp{0.5cm}

In questo modo, dal raffronto dell'efficienza del Trigger neutro in tali
condizioni di utilizzo con l'efficienza in assenza di accidentali, e
dalla conoscenza del rate di decadimenti \ks e \kl e del rate di eventi
accidentali, e' possibile stimare l'efficienza dell'apparato di trigger
in condizioni reali di funzionamento.

Il rapporto $out0/in1$ indica la percentuale di eventi buoni che, a
causa di un gamma accidentale, non vengono accettati dal trigger.
Il rapporto $out1/in0$ evidenzia la percentuale di eventi non buoni in
ingresso che vengono invece giudicati buoni dal trigger in presenza di
un evento accidentale; uno dei possibili casi in cui si puo' verificare
tale situazione e', ad esempio, un decadimento in \dpiz con soli tre
gamma sul calorimetro elettromagnetico ($in0$) piu' un quarto gamma
accidentale che apporta una correzione alla ricostruzione di energia,
baricentro dell'energia e vertice di decadimento, altrimenti scartati,
tale da far accettare l'evento come decadimento in \dpiz.

Detto $r_d$ il rate di decadimenti \ks e \kl, e $r_a$ il rate di eventi
accidentali sul calorimetro elettromagnetico, la percentuale di eventi
buoni contaminati da un evento accidentale e' data da:
\beq P=\frac{1}{r_d} (2\tau r_d\cdot r_a)=2\tau r_a \eeq
dove $\tau$ e' la larghezza temporale del segnale di shaper.

Dalla conoscenza quindi dell'efficienza $e_0$ in assenza di eventi
esterni, dell'efficienza in presenza di eventi accidentali come
precedentemente definita, l'efficienza in condizioni reali di
funzionamento $e$ del Trigger neutro nell'acquisizione di eventi buoni
risulta essere:
\beq e=e_0-p_{10}+p_{01} \eeq
dove $p_{10}$ e' la percentuale di eventi buoni sporcati da eventi
accidentali e non accettati dal trigger, e $p_{01}$ e' la percentuale
di eventi non buoni resi accettabili dal gamma accidentale.\\
I valori di $p_{10}$ e $p_{01}$ risultano dati da:
\beq p_{10}=P\cdot \frac{out0}{in1} \eeq
\beq p_{01}=P\cdot \frac{out1}{in0} \eeq
dove i rapporti relativi alle inefficienza dell'apparato si riferiscono
ai valori calcolati in presenza di accidentali con rate equivalente al
rate degli eventi di decadimento.

In condizioni di rate di decadimenti di \ks e \kl pari a $1\s MHz$, e
rate di eventi accidentali pari a $10\s KHz$,
la percentuale di eventi contaminati da un gamma accidentale e' pari a
\beq P=2.2\cdot10^{-3} \eeq
ovvero la probabilita' di contaminazione di un evento da parte di un
accidentale e' tra\-scu\-ra\-bi\-le ai fini dell'esperimento.

Applicando i tagli sulle quantita' fisiche ricostruite precedentemente
introdotti (tagli modificati), si ottiene una efficienza di
funzionamento dell'apparato di trigger, calcolata come rapporto tra gli
eventi giudicati buoni in uscita e gli eventi effettivamente buoni in
ingresso:\\
\centerline{\bksdpiz:\hsp{0.5cm} $e=0.754$}
\centerline{\bkldpiz:\hsp{0.5cm} $e=0.384$}
Questo parametro indica l'effettiva efficienza dell'apparato di trigger
in presenza di tali tagli, senza tenere conto della possibile
dispersione dei gamma, e quindi del fatto che molti eventi giungono con
un numero inferiore di gamma rispetto a quelli prodotti sul
calorimetro.\\
Per quanto riguarda invece i decadimenti di tipo \kltpiz, una
percentuale di eventi pari a $6.09\%$ viene candidata dal trigger
neutro ad essere frutto di un decadimento in \dpiz.


\end{document}